# Towards a Switching-Algebraic Theory of Weighted Monotone Voting Systems: The Case of Banzhaf Voting Indices

[1]**Ali Muhammad Rushdi** and [2,3]**Muhammad Ali Rushdi**

[1]Department of Electrical and Computer Engineering, Faculty of Engineering,
King Abdulaziz University, P. O. Box 80200, Jeddah, 21589, Kingdom of Saudi Arabia
{arushdi@kau.edu.sa; arushdi@ieee.org}

[2]Department of Biomedical and Systems Engineering, Faculty of Engineering,
Cairo University, Giza 12613, Arab Republic of Egypt
{mrushdi@eng1.cu.edu.eg}
[3]School of Information Technology,
New Giza University, Giza 12256, Arab Republic of Egypt
{Muhammad.Rushdi@ngu.edu.eg}

**Abstract**: This paper provides a serious attempt towards constructing a switching-algebraic theory for weighted monotone voting systems, whether they are scalar-weighted or vector-weighted. The paper concentrates on the computation of a prominent index of voting powers, *viz.*, the Banzhaf voting index. This computation involves two distinct operations: (a) either Boolean differencing (Boolean differentiation) or Boolean quotient construction (Boolean restriction), and (b) computation of the weight (the number of true vectors or minterms) of a switching function. We introduce novel Boolean-based symmetry-aware techniques for computing the Banzhaf index by way of four voting systems. The first system is a small family bi-cameral parliament modeled by a vector-weighted voting system. The second system models the United Nations Security Council, which has members with veto-power, as a scalar-weighted voting system. Two versions (an extended version and a reduced one) are given for the third system, which is a scalar-weighted voting system that represents the Scottish Parliament of 2007. Solution of the extended version is obtained incrementally from that of the reduced one. Both versions demonstrate that the Banzhaf index might be computed in terms of the decision function of the system or in terms of its complement. The fourth system is a tri-cameral parliament modeled by a vector-weighted voting system of dimension 3. Each chamber of this parliament is a specific k-out-of-n system. Most powerful among members of this parliament are those of the chamber whose k-out-of-n system possesses the narrowest region of useful redundancy. The paper finally outlines further steps needed towards the establishment of a full-fledged switching-algebraic theory of weighted monotone voting systems. Througout the paper, a tutorial flavour is retained, multiple solutions of consistent results are given, and a liasion is established among game-theoretic voting theory, switching algebra, and sytem reliability analysis.

## 1. Introduction

A voting system is a set of rules that specifies how voters choose among several candidates or alternatives [1-5]. A yes-no voting system is a voting system where the choice is between two alternatives. Typically, there is a simple question of adopting a potentially forthcoming alternative (a bill, resolution or amendment) against the status quo, which stands as an already existing alternative [1-5]. In a switching-algebraic context [6-18], the decision taken by the voting system is represented by a switching (two-valued Boolean) function $f(\boldsymbol{X})$, such that $f(\boldsymbol{X}) = 1$ if the decision is upheld, and $f(\boldsymbol{X}) = 0$ if it is rejected. Here, the binary vector $\boldsymbol{X} = [X_1 \ X_2 \ \ldots \ X_n]^T$ is an n-tuple of the votes $X_i$ [$1 \leq i \leq n$] cast by voters, where $X_i$ is 1 or 0 if voter $i$ says 'yes' or 'no', respectively. In the sequel, we use the term of a voting system to implicitly mean a monotone voting system [2], i.e., one for which the decision function $f(\boldsymbol{X})$ is a semi-coherent switching function [10, 15-21]. This is a monotonically non-decreasing (or simply a monotonically increasing function) [22], for which $f(\boldsymbol{X}) \geq f(\mathbf{Y})$ for $\boldsymbol{X} \geq (\mathbf{Y})$. Since the property of causality ($\{f(\mathbf{0}) = 0\} \wedge \{f(\mathbf{1}) = 1\}$) is implied by that of monotonicity (for non-constant functions representing non-fictitious systems), then a semi-coherent $f(\mathbf{X})$ (and hence, a monotone voting system) possesses the property of causality as well.

A coalition is any set of yes-voters [2]. If the debated alternative is upheld (i.e., if the bill, resolution or amendment is passed) when supported merely by a certain coalition, this coalition is viewed as a winning one. Otherwise, it is considered a losing one. The coalition is empty if no voter belongs to it and it is the grand coalition if all voters belong to it [2, 17, 23, 24]. By contrast, a primitive coalition is a specification of the status of all voters [8, 10, 11, 16, 17]. This can be a primitive winning coalition (PWC), that corresponds to a true vector (minterm) of $f(\boldsymbol{X})$, or a primitive losing coalition (PLC), that corresponds to a false vector (minterm) of $f(\boldsymbol{X})$. The concept of a primitive coalition is convenient for Boolean-based analysis, but it is alien to voting theory. It is typically a mixture of yes-voters and no-voters (typically expressed by a product of uncomplemented literals and complemented ones). By contrast, the concept of a coalition is popular in voting theory, and it concerns solely yes-voters. For a voting system of $n$ voters, there are $2^n$ coalitions (including the empty one) and there are also $2^n$ primitive coalitions. If the uncomplemented literals representing voters that are present in a minimal winning coalition ($MWC$) are ANDed, they constitute a product that is a prime implicant of $f(\boldsymbol{X})$, and if the complemented literals representing voters that are absent in a maximal losing coalition ($MLC$) are ANDed, they constitute a product that is a prime implicant of its complement $\overline{f}(\boldsymbol{X})$. In a monotone voting system, the *empty* coalition is losing ($f(\mathbf{0}) = 0$), while the *grand* coalition is winning ($f(\mathbf{1}) = 1$) [2, 17], and hence the cardinality of a $MWC$ is at least 1, while that of a maximal losing one is at most $(n - 1)$.

By definition, a yes–no voting system is scalar-weighted (or simply weighted) if it can be described by specifying a weight for each voter and a quota so that the winning coalitions are precisely the ones whose total weight meets or exceeds the quota. Such a definition does not address situations involving veto power. However, if a scalar-weighted voting system is altered by giving some of the voters the veto power (the power of stopping a motion from passing), the resulting yes–no voting system is again a scalar-weighted voting system (albeit with alternative weights and threshold). Examples of simple scalar-weighted voting systems (ones without veto power) include the voting systems of the European Economic Community, the European Union before the eastern extension in 2004, the US Electoral College, the Board of Governors of the International Monetary Fund, shareholder elections in a company, and the administrative councils of many companies and banks. Examples of non-simple scalar-weighted voting systems (ones modelling systems in which the veto power is accounted for) include the voting system of the United Nations Security Council (UNSC) as well as the legislative processes in many democracies.

A scalar-weighted voting system is sometimes described as a scheme in which one voter possesses $W_n$ votes, as opposed to the more basic scheme of one voter with one vote. It is usually depicted mathematically as $(T; W_1, W_2, \dots, W_n)$, where $T$ denotes its threshold or quota, and $W_1, W_2, \dots, W_n$ depict the weights of tits constituent voters. Both the quota $T$ and the weights $W_j$ are strictly positive for a monotone system. The quota cannot be more than the sum of the weights (or else no motion/resolution would ever be passed), and it is prudent that the quota be strictly greater than half of the sum of the weights (or else too few motions/resolutions would be rejected, and more seriously, two disjoint coalitions could be simultaneously winning). A scalar-weighted system is described herein by a switching (two-valued Boolean) threshold function $f(\boldsymbol{X})$, such that is $f(\boldsymbol{X}) = 1$ if the resolution considered is passed and $f(\boldsymbol{X}) = 0$ if the resolution is rejected. A threshold switching function $f(\mathbf{X})$ of $n$ variables is characterized by $(n+1)$ (rather than $2^n$) coefficients, namely a threshold $T$ and weights $\boldsymbol{W} = [W_1, W_2, \dots, W_n]^{\mathrm{T}}$, such that [6-11, 15-18, 25-28]

$$f(\mathbf{X}) = 1 \ \text{ iff } \ F(\mathbf{X}) \equiv \sum_{i=1}^{n} W_i X_i \geq T. \tag{1}$$

The inequality in (1) is designated a knapsack cover constraint when it has strictly positive weights and threshold, in which case $f(\mathbf{X})$ is termed a semi-coherent or monotone threshold switching function. Such a function might be described as scale-invariant [29], since multiplying every weight and the threshold by the same positive constant does not change the function. If further, the function $f(\mathbf{X})$ is genuine (non-vacuous) in each of its $n$ arguments (i.e., if each of the arguments is relevant and not dummy), then $f(\mathbf{X})$ is called (fully) coherent. Unfortunately, the subtle distinction between the terms coherent and semi-coherent is blurred in the literature. A coherent threshold switching function is a natural description for the success of a coherent/semi-coherent threshold reliability system [6, 7, 15], or equivalently, for the decision made by a monotone scalar-weighted voting system [8, 10,11, 16, 17].

A necessary (but not sufficient) condition for a yes-no voting system to be scalar-weighted is that it must be swap robust [1, 2], i.e., any swap or one-for-one exchange of two distinct voters between two winning coalitions leaves at least one of the resulting two coalitions winning. In the sequel, we will present two voting systems that are not swap robust and hence not scalar-weighted. For these systems, we can generalize the notions of ‘threshold’ and ‘weights’ from scalar numbers to vectors, and then these systems (and all yes–no voting systems) will be referred to henceforth as ‘vector-weighted.’ We will use the designation ‘weighted-voting system’ to refer to both scalar-weighted and vector-weighted ones, and hence employ this designation as an umbrella for all yes-no voting systems. Taylor and Pacelli [2] introduced the notion of the dimension of a yes–no voting system as the minimum number of scalar-weighted voting systems needed to realize the given system as their intersection. However, they reported no system of a dimension beyond two, and explicitly stated that they knew of no real-world voting system of dimension 3 (or more). Later, Freixas [30] asserted that the voting system of the European Union enlargement according to the Nice Treaty (under the two different decision rules of simple majority or qualified two-thirds majority) has exactly dimension 3. Also, Cheung and Ng [31] stated that the voting system used in the Hong Kong Legislative Council is a three-dimensional one. Kurz and Napel [32] proved that the Lisbon voting system of the Council of the European Union (EU) (effective 2014-2020) cannot be represented as the intersection of six or fewer weighted games, i.e., its dimension $d$ (to be exactly determined yet) is at least 7 (actually, $7 \leq d \leq 13, 368$). With this proof, they set with their lower bound of 7 a new record for the dimension of a real-world voting body, and they posed the determination of the aforementioned dimension $d$ as a challenge. In response to this challenge, Chen et *al.* [33] obtained an upper bound for the dimension $d$ of the EU voting system by showing it is at most 25, while Kober and Weltge [34] provided its first improved lower bound by showing that it is at least 8. Therefore, the strongest currently known bounds on d are $8 \leq d \leq 25$. In the sequel, we will present a hypothetical voting system of dimension 3 (that could be indefinitely extended), and consider its reduction to one of dimension 2.

Many indices of voting power are in use nowadays. Two of these are effectively the mainstream standard ones. These are the Shapley–Shubik power index [2, 11, 35-37] and the Penrose–Banzhaf–Coleman index (referred to herein as the Banzhaf index [2, 10, 14, 16, 36-39] for short), which was introduced by Banzhaf [38], with a pre-contribution by Penrose [40], and a post-contribution by Coleman [41]. These two indices are based on permutational and coalitional/combinatorial considerations, respectively. The Shapley–Shubik index is the one that appeared earlier and that attracted more widespread attention. Other notable indices for voting power are the Johnston index [2, 42, 43], the Deegan-Packel index [2, 43, 44], the Member Bargaining Power (MPP) measure [45, 46], and the Public Good Index (PGI), introduced by Holler [47] and axiomatized by Holler and Packel [48]. Many useful comparisons between the various voting-power indices abound in the open literature [36, 43, 49-51].

Our main concern in this paper is the switching-algebraic computation of the Banzhaf index or Banzhaf voting power of an individual voter $P$ in a voting system. This voting power is typically called the total Banzhaf power $TBP(P)$ of voter $P$ [2]. It is the number of times voter $P$ is decisive, i.e., the number of winning states or configurations in which this voter is among the proponents of the resolution (proposal) such that a switch of the voter to join the opponents changes the system state from one of winning to that of losing. As stated earlier, each of the $2^n$ states or configurations of an n-member weighted voting system is called a primitive coalition [8, 10, 11, 16, 17]. The Total Banzhaf Power $TBP(P)$ is then the number of primitive winning coalitions ($PWCs$) of which voter $P$ is a member such that defection of $P$ causes a transition or swing to a primitive losing coalition ($PLC$). Implicit in the definition of the Banzhaf index is the assumption that system states or configurations (primitive coalitions) are equally likely. This in turn necessitates that voters cast their votes independently of each other.

The organization of the rest of this paper is as follows. Section 2 deals with the switching-algebraic computation of Banzhaf indices. This section has four subsections. Subsection 2.1 presents several switching-algebraic definitions of the Banzhaf index for monotone systems, with some definitions making explicit utilization of the nature of positively-polarized unate functions. Subsection 2.2 defines and presents pertinent formulas for the Boolean difference (Boolean derivative). Subsection 2.3 explores the computation of the weight of a switching function. Subsection 2.4 introduces symmetric switching functions and their Banzhaf indices. Sections 3 to 6 illustrate novel Boolean-based symmetry-aware techniques for computing the Banzhaf index by way of four voting systems. The first system (Section 3) is a small family bi-cameral parliament modeled by a vector-weighted voting system of dimension 2. The second system (Section 4) models the United Nations Security Council. which has members with veto-power, as a scalar-weighted voting system. Section 5 presents two versions (an extended version and a reduced one) of a scalar-weighted voting system that represents the Scottish Parliament of 2007. Solution of the extended version is obtained incrementally from that of the reduced one. Both versions demonstrate that the Banzhaf index might be computed in terms of the function $f(\boldsymbol{X})$ or in terms of its complement $\overline{f}(\boldsymbol{X})$. The fourth system (Section 6) is a tri-cameral parliament modeled by a vector-weighted voting system of dimension 3. Each chamber of this parliament is a specific k-out-of-n system. Most powerful among members of this parliament are those of the chamber whose k-out-of-n system possesses the narrowest region of useful redundancy. Section 7 concludes the paper. To make the paper self-contained, its main text is supplemented with four appendices. Appendix A is a brief introduction to unate switching functions, while Appendix B proves that two specific definitions of the Banzhaf index in subsection 2.1 are equivalent. Appendix C demonstrates how the cumulative combinatorial coefficient can be efficiently computed. Finally, Appendix D shows that disjointness (of products in a sum-of-products form) is preserved when a Boolean quotient is constructed, but not when a partial Boolean derivative is obtained.

## 2. Switching-Algebraic Computation of Banzhaf Indices

### 2.1. Switching-algebraic definition of Banzhaf indices

The total Banzhaf power of voter number $m$ is defined for any monotone voting system as [7, 10, 14, 16, 17, 39, 52]

$$TBP(X_m) = wt\left(\frac{\partial f(\boldsymbol{X})}{\partial X_m}\right) = wt((f(\boldsymbol{X})/X_m) \oplus (f(\boldsymbol{X})/\overline{X}_m)), \quad (1 \le m \le n). \tag{2}$$

Here, the symbol $\frac{\partial f(\boldsymbol{X})}{\partial X_m}$ denotes the partial derivative of the voting system Boolean function $f(\boldsymbol{X})$ w.r.t. its argument $X_m$ [53] (See subsection 2.2), while the symbol $wt(\ldots)$ denotes the weight or number of true vectors of a switching function [54-61] (See subsection 2.3). The ratios $(f(\boldsymbol{X})/X_m)$ and $(f(\boldsymbol{X})/\overline{X}_m)$ denote the Boolean quotients [62-67] of $f(\boldsymbol{X})$ w.r.t. the literals $X_m$ and $\overline{X}_m$, respectively, i.e.

$$f(\boldsymbol{X})/X_m = f(\boldsymbol{X}|(X_m = 1). \tag{3}$$

$$f(\boldsymbol{X})/\overline{X}_m = f(\boldsymbol{X}|(X_m = 0). \tag{4}$$

For a monotone system, the function $f(\boldsymbol{X})$ is a positively-polarized unate function (See Appendix A), and hence $TBP(X_m)$ might be computed alternatively as the number of primitive winning coalitions ($PWCs$) in which the variable $X_m$ appears as uncomplemented literal (i.e., just $X_m$) minus those in which it appears as the complemented one (i.e., as $\overline{X}_m$). This is the same as double the number of $PWCs$ in which $X_m$ is present as a proponent minus the total number of $PWCs$ [2]. Mathematically, this can be stated for $(1 \le m \le n)$ as

$$TBP(X_m) = 2\,wt((f(\boldsymbol{X})/X_m)X_m) - wt\big(f(\boldsymbol{X})\big) = 2\,wt(f(\boldsymbol{X})/X_m)\,wt(X_m) - wt\big(f(\boldsymbol{X})\big)$$
$$= 2\,wt(f(\boldsymbol{X})/X_m) - wt\big(f(\boldsymbol{X})\big). \qquad (5)$$

Appendix B provides a proof that expression (5) is equivalent to definition (2) when $f(\boldsymbol{X})$ is a unate function of positive polarity. Appendix B also adds another expression of $TBP(X_m)$ for $(1 \le m \le n)$ under the same conditions

$$TBP(X_m) = wt\big(f(\boldsymbol{X})\big) - 2\,wt((f(\boldsymbol{X})/\overline{X}_m)\overline{X}_m) = wt\big(f(\boldsymbol{X})\big) - 2\,wt(f(\boldsymbol{X})/\overline{X}_m). \qquad (6)$$

If we add equation (5) to equation (6) and then divide by 2, we obtain yet another formula for $TBP(X_m)$, namely

$$TBP(X_m) = wt((f(\boldsymbol{X})/X_m)X_m) - wt((f(\boldsymbol{X})/\overline{X}_m)\overline{X}_m) = wt(f(\boldsymbol{X})/X_m) - wt(f(\boldsymbol{X})/\overline{X}_m). \qquad (7)$$

The weight $wt\big(f(\boldsymbol{X})\big)$ of the function $f(\boldsymbol{X})$ is given through its Boole-Shannon expansion by

$$wt\,\big(f(\boldsymbol{X})\big) = wt\,\Big((f(\boldsymbol{X})/X_m)\,X_m \vee \big(f(\boldsymbol{X})/\,\overline{X}_m\big)\,\overline{X}_m\Big) = wt(\,(f(\boldsymbol{X})/X_m)\,X_m \oplus (f(\boldsymbol{X})/\,\overline{X}_m)\,\overline{X}_m) = wt(f(\boldsymbol{X})/X_m)wt(\,X_m) + wt\big(f(\boldsymbol{X})/\,\overline{X}_m\big)wt(\,\overline{X}_m) = wt(f(\boldsymbol{X})/X_m) + wt\big(f(\boldsymbol{X})/\,\overline{X}_m\big). \qquad (8)$$

It is worth noting that the expressions in (2) and (8) are not the same, since the Boolean quotients $f(\boldsymbol{X})/X_m$ and $f(\boldsymbol{X})/\,\overline{X}_m$ in (2) are not disjoint, while the terms $(f(\boldsymbol{X})/X_m)\,X_m$ and $(f(\boldsymbol{X})/\,\overline{X}_m)\,\overline{X}_m$ in (8) are disjoint.

Typically, the raw total Banzhaf powers are each divided by $2^{n-1}$ (the maximum possible weight for $\frac{\partial f(\boldsymbol{X})}{\partial X_m}$), which allows a certain probability interpretation. Alternatively, these raw powers are normalized by dividing each of the raw values $TBP(X_m)$ by the sum of these values, which allows the resulting normalized indices to have a different (albeit criticized) probability interpretation, and also allows easy comparison with the other prominent types of indices, *viz.*, the Shapley-Shubik indices [36]. Therefore, the normalized total Banzhaf power is given by

$$NTBP(X_m) = TBP(X_m)/\sum_{k=1}^{n} TBP(X_k), \qquad (1 \le m \le n). \qquad (9)$$

Our present aim is to explore the calculation of the Banzhaf power index from the perspective of switching algebra, rather than from the traditional paradigm of game theory. The technique to be proposed herein is a switching-algebraic technique that is basically an enumeration method. This technique tries to make the most of recent developments in switching theory [22, 53, 55, 64, 68-70] as well as of symmetry features that are inherent to many voting systems.

### 2.2. The Boolean difference or derivative

A switching function (a two-valued Boolean function) of *n* variables is a mapping from $B_2^n = \{0,1\}^n$ into $B_2 = \{0,1\}$, that is denoted by $f(\boldsymbol{X}) = f(X_1, X_2, \cdots, X_m, \ldots, X_{n-1}, X_n)$. The partial derivative (or Boolean difference) of $f(\boldsymbol{X})$ w.r.t. $X_m$ $(1 \le m \le n)$ is [7, 8, 10, 16, 17, 22, 26, 52, 53]

$$\frac{\partial f}{\partial X_m} = f\big(X_1, X_2, \cdots, \overline{X}_m, \ldots, X_{n-1}, X_n\big) \oplus f(X_1, X_2, \cdots, X_m, \ldots, X_{n-1}, X_n). \qquad (10)$$

This can be seen to be equivalent to [53]

$$\frac{\partial f}{\partial X_m} = f(\boldsymbol{X}|X_m = 0) \oplus f(\boldsymbol{X}|X_m = 1). \qquad (11)$$

where

$$f(\boldsymbol{X}|X_m = 0) = f(X_1, X_2, \cdots, 0, \dots, X_{n-1}, X_n) = f(\boldsymbol{X}) \,/\, \overline{X}_m, \qquad (12a)$$

$$f(\boldsymbol{X}|X_m = 1) = f(X_1, X_2, \cdots, 1, \dots, X_{n-1}, X_n) = f(\boldsymbol{X}) \,/\, X_m, \qquad (12b)$$

are called subfunctions, restrictions or quotients (ratios) of $f(\boldsymbol{X})$. Their Karnaugh maps are obtained by *splitting* the Karnaugh map of $f(\boldsymbol{X})$ into two halves, *viz.*, the asserted domains for $\overline{X}_m$ and $X_m$. The Boolean difference is then obtained by *folding* one of these two halves onto the other and performing *XORing* cell-wise [53]. Some of the important properties of the Boolean difference are (for $A$ and $B$ being independent of $X_m$ ) [22, 53]

$$\{f(\boldsymbol{X}) = f_1(\boldsymbol{X}) \oplus f_2(\boldsymbol{X})\} \quad \longrightarrow \quad \{\frac{\partial f(\boldsymbol{X})}{\partial X_m} = \frac{\partial f_1(\boldsymbol{X})}{\partial X_m} \oplus \frac{\partial f_2(\boldsymbol{X})}{\partial X_m}\}, \qquad (13)$$

$$\{f(\boldsymbol{X}) = f_1(\boldsymbol{X}) \vee f_2(\boldsymbol{X})\} \quad \longrightarrow \quad \{\frac{\partial f(\boldsymbol{X})}{\partial X_m} = \overline{f}_1(\boldsymbol{X}) \frac{\partial f_2(\boldsymbol{X})}{\partial X_m} \oplus \frac{\partial f_1(\boldsymbol{X})}{\partial m} \overline{f}_2(\boldsymbol{X}) \oplus \frac{\partial f_1(\boldsymbol{X})}{\partial X_m} \frac{\partial f_2(\boldsymbol{X})}{\partial X_m}\}, \qquad (14)$$

$$\frac{\partial f}{\partial X_m} = \frac{\partial f}{\partial \overline{X}_m} = \frac{\partial \overline{f}}{\partial X_m} = \frac{\partial \overline{f}}{\partial \overline{X}_m}, \qquad (15)$$

$$\frac{\partial (AX_m)}{\partial X_m} = \frac{\partial (AX_m)}{\partial \overline{X}_m} = A, \qquad (16)$$

$$\frac{\partial (B)}{\partial X_m} = \frac{\partial (B)}{\partial \overline{X}_m} = 0. \qquad (17)$$

Equation (13) indicates that the partial differentiation operator ($\frac{\partial}{\partial X_i}$) commutes with the XOR operator. By contrast, the partial differentiation operator does not commute with the OR operator, and a quite involved formula is needed (14) to differentiate an ORed expression. That is the reason why we prefer to pre-process a sum-of-products expression by rendering it in a disjoint form so as to be able to legitimately replace OR operators by XOR ones. Equation (15) suggests that $TBP(X_m)$ could be computed as the weight of $\frac{\partial \overline{f}(\boldsymbol{X})}{\partial X_m}$ or $\frac{\partial \overline{f}(\boldsymbol{X})}{\partial \overline{X}_m}$ but frequently $\overline{f}(\boldsymbol{X})$ is more complicated than $f(\boldsymbol{X})$. Usually, a large voting system has minimal winning coalitions ($MWCs$) that are much fewer than its minimal losing coalitions ($MLCs$).

### 2.3. Computing the weight of a switching function

We interpret a switching function $f(\boldsymbol{X}) = f(X_1, X_2, \cdots, X_{n-1}, X_n)$ as the output column of its truth table $\boldsymbol{f}$, i.e., a binary vector of length $2^n$, namely

$$\boldsymbol{f} = [f(0,0,\dots 0,0) \quad f(0,0,\dots,0,1) \quad f(0,0,..,1,0) \ \dots \ f(1,1,..,1,1)]^T. \qquad (18)$$

The weight of the switching function $f(\boldsymbol{X})$ is the number of ones in its truth-table vector $\boldsymbol{f}$. This is denoted by $wt(f)$, and is bounded by $0 \leq wt(f) \leq 2^n$ [54-61]. An $n$-variable function $f$ is said to be balanced if its output column in the truth table contains equal numbers of *0's* and *1's* ($i.e., wt(f) = 2^{n-1}$). Obviously, if a switching function $f(\boldsymbol{X})$ is expanded about $m$ of its arguments into $2^m$ subfunctions, then the weight of the parent function $f(\boldsymbol{X})$ is the sum of the weights of these subfunctions.

If the function $f(\boldsymbol{X})$ is available in a general s-o-p form that is not necessarily disjoint,

$$f(\boldsymbol{X}) = \bigvee_{\mathrm{i}=1}^{\mathrm{n_p}} \mathrm{P_i}, \tag{19}$$

then the weight of $f(\boldsymbol{X})$ is given by an appropriate version of the Inclusion-Exclusion (IE) Principle [67, 71-76] as follows

$$wt(f) = \sum_{\mathrm{i}=1}^{\mathrm{n_p}} wt\{\mathrm{P_i}\} - \sum\sum_{1\le \mathrm{i}<j\le \mathrm{n_p}} wt\{\mathrm{P_i} \wedge \mathrm{P_j}\} + \sum\sum\sum_{1\le \mathrm{i}<j<k\le \mathrm{n_p}} wt\{\mathrm{P_i} \wedge \mathrm{P_j} \wedge \mathrm{P_k}\} - \ldots + (-1)^{\mathrm{n_p}-1}\, wt\{\bigwedge_{\mathrm{i}=1}^{\mathrm{n_p}} \mathrm{P_i}\}. \tag{20}$$

where the weight of a product $P_k$ is equal to $2^{(n-\ell(P_k))}$, i.e. to 2 raised to a power of the total number of variables minus the number of irredundant literals in the product. The symbol $\ell(P_k)$ denotes the number of irredundant literals in the product $P_k$, e.g. $\ell(1) = 0$, $\ell(X_i) = \ell(\overline{X}_i) = 1$, $\ell(X_i X_j) = \ell(X_i \overline{X}_j) = 2$. The weight $wt(P_k)$ of a product $P_k$ depends not only on its number of irredundant literals $\ell(P_k)$, but also on the total number $n$ of variables involved. In fact, the weight $wt(P_k)$ is equal to 2 raised to a power of the total number of missing variables $(n - \ell(P_k))$. The logical value 0 is not considered a product $D_k$. But anyhow we assume that $\ell(0)$ tends to infinity ($\infty$), so were we to have $P_k = 0$, it would contribute nothing to $wt(f)$. Here, we assume that none of the products $P_k$ has any redundant literal (redundant literals, if any, can be eliminated via idempotency of the AND operator ($X_i \wedge X_i = X_i$ *and* $\overline{X}_i \wedge \overline{X}_i = \overline{X}_i$)).

If the switching function $f(\boldsymbol{X})$ be expressed by the disjoint sum-of-products (s-o-p) form [64, 67, 71, 74, 77-79]

$$f(\boldsymbol{X}) = \bigvee_{k=1}^{m} D_k, \tag{21}$$

where

$$D_i \wedge D_j = 0, \qquad \forall\ \mathrm{i, j}, \tag{22}$$

$$D_k = \left(\bigwedge_{i\in I_{k_1}} X_i\right)\left(\bigwedge_{i\in I_{k_2}} \overline{X}_i\right), \qquad \forall\ \mathrm{k}, \tag{23}$$

Here, $I_{k_1}$ and $I_{k_2}$ are the sets of indices for uncomplemented literals and complemented literals in the product $D_k$, and $\ell(D_k)$ is the sum of cardinalities of the sets $I_{k_1}$ and $I_{k_2}$. Then, we obtain the weight of $f(\boldsymbol{X})$ as

$$wt(f) = \sum_{k=1}^{m} wt(D_k) = \sum_{k=1}^{m} 2^{(n-\ell(D_k))}. \tag{24}$$

If the function $f(\boldsymbol{X})$ is a symmetric switching function $Sy(n;\boldsymbol{A};\boldsymbol{X})$, then its weight can be obtained by summing the combinatorial (binomial) coefficients $n$ choose $a$, denoted $c(n,a)$, for all integers $a$ that belong to the characteristic set $\boldsymbol{A}$, namely:

$$wt(Sy(n;\boldsymbol{A};\boldsymbol{X})) = \sum_{a\in\boldsymbol{A}} c(n,a). \tag{25}$$

If a function $f(\boldsymbol{X},\boldsymbol{Y})$ is a conjunction of two functions $f_1(\boldsymbol{X})$ and $f_2(\boldsymbol{Y})$, where $\boldsymbol{X}$ and $\boldsymbol{Y}$ are non-overlapping sets of arguments, then the weight of $f(\boldsymbol{X},\boldsymbol{Y})$ is the product of the weights of $f_1(\boldsymbol{X})$ and $f_2(\boldsymbol{Y})$.

$$\{f(\boldsymbol{X},\boldsymbol{Y}) = f_1(\boldsymbol{X}) \wedge f_2(\boldsymbol{Y})\} \ \longrightarrow \ \{wt(f(\boldsymbol{X},\boldsymbol{Y})) = wt(f_1(\boldsymbol{X})) * wt(f_2(\boldsymbol{Y}))\}. \tag{26}$$

If a function $f(\boldsymbol{X})$ is a disjunction of two disjoint functions $f_1(\boldsymbol{X})$ and $f_2(\boldsymbol{X})$, then its weight is the sum of their weights

$$\{f(\boldsymbol{X}) = f_1(\boldsymbol{X}) \vee f_2(\boldsymbol{X}) = f_1(\boldsymbol{X}) \oplus f_2(\boldsymbol{X}), \quad f_1(\boldsymbol{X}) \wedge f_2(\boldsymbol{X}) = 0\} \ \longrightarrow$$
$$\{wt(f(\boldsymbol{X})) = wt(f_1(\boldsymbol{X})) + wt(f_2(\boldsymbol{X}))\}. \tag{27a}$$

If the functions $f_1(\boldsymbol{X})$ and $f_2(\boldsymbol{X})$ are not disjoint, then

$$wt(f_1(\boldsymbol{X}) \vee f_2(\boldsymbol{X})) = wt(f_1(\boldsymbol{X})) + wt(\, f_2(\boldsymbol{X})) - wt(f_1(\boldsymbol{X}) \wedge f_2(\boldsymbol{X})), \tag{27b}$$

$$wt(f_1(\boldsymbol{X}) \oplus f_2(\boldsymbol{X})) = wt(f_1(\boldsymbol{X})) + wt(\, f_2(\boldsymbol{X})) - 2\, wt(f_1(\boldsymbol{X}) \wedge f_2(\boldsymbol{X})). \tag{27c}$$

The weights of a function (of $n$ arguments) and its complement add to $2^n$

$$wt\left(\overline{f}(\boldsymbol{X})\right) = wt(1 \ \oplus \ f(\boldsymbol{X})) = 2^n - wt(f(\boldsymbol{X})). \tag{28}$$

## 2.4. Banzhaf indices for symmetric switching functions

A symmetric switching function (SSF) is a two-valued Boolean function defined [17, 18, 64, 66, 80, 81] as

$$f(\boldsymbol{X}) = Sy(n;\boldsymbol{A};\boldsymbol{X}) = \ Sy(\,n;\{a_1,a_2,\ldots,a_m\};\ X_1,X_2,\ldots,X_n), \tag{29}$$

and is specified *via* its number of inputs $n$, its inputs $\boldsymbol{X} = \ [X_1,X_2,\ldots,X_n]^{\mathrm{T}}$, and its characteristic set

$$\boldsymbol{A} = \ \{a_0,a_1,\ldots,a_m\} \ \subseteq \ \boldsymbol{I}_{n+1} \ = \{0,1,2,\ \ldots,n\},\ \{m \ \le n\}. \tag{30}$$

This function has the value 1 if and only if the arithmetic sum $\sum_{j=1}^{n} X_j$ belongs to the characteristic set $\boldsymbol{A}$, and has the value 0, otherwise. The complement $\bar{f}$ of the above SSF has a characteristic set $\overline{\boldsymbol{A}}$ defined by the complementary set *w.r.t.* $\boldsymbol{I}_{n+1} = \ \{0,1,2,\ \ldots,n\}$, given by the set difference $(\boldsymbol{I}_{n+1}/\boldsymbol{A})$, also denoted as $(\boldsymbol{I}_{n+1} - \boldsymbol{A})$, or

$$\overline{\boldsymbol{A}} \ = \ \{0,1,2,\ \ldots,n\} - \ \{a_0,a_1,\ldots,a_m\} \tag{31}$$

and hence, $\bar{f}$ can be expressed as

$$\bar{f} \ \ = Sy(n;\overline{\boldsymbol{A}};\boldsymbol{X}). \tag{32}$$

The *ANDing*, *ORing*, and *XORing* of two SSFs of the same arguments $\boldsymbol{X}$, and of characteristic sets $\boldsymbol{A_1}$ and $\boldsymbol{A_2}$ are SSFs with characteristic sets equal to the intersection, the union, and the symmetric difference of $\boldsymbol{A_1}$ and $\boldsymbol{A_2}$, respectively, *i.e.*,

$$Sy(n;\boldsymbol{A_1};\boldsymbol{X}) \ \ \wedge \ \ Sy(n;\boldsymbol{A_2};\boldsymbol{X}) \ \ = \ \ Sy(n;\boldsymbol{A_1} \cap \ \boldsymbol{A_2};\boldsymbol{X}), \tag{33a}$$

$$Sy(n;\boldsymbol{A_1};\boldsymbol{X}) \quad \vee \quad Sy(n;\boldsymbol{A_2};\boldsymbol{X}) \; = \; Sy(n;\boldsymbol{A_1} \cup \boldsymbol{A_2};\boldsymbol{X}), \tag{33b}$$

$$Sy(n;\boldsymbol{A_1};\boldsymbol{X}) \quad \oplus \quad Sy(n;\boldsymbol{A_2};\boldsymbol{X}) = \; Sy(n;\boldsymbol{A_1} \oplus \boldsymbol{A_2};\boldsymbol{X}). \tag{33c}$$

The *ANDing* of two SSFs of the same arguments $\boldsymbol{X}$, and of disjoint characteristic sets is $0$, and hence these two functions are disjoint. The Boole-Shannon expansion for the SSF $f$ in (29) about any of its variables $X_m$ $(1 \leq m \leq n)$ can be stated as follows [64, 66, 67, 80, 81]

$$Sy(n;\boldsymbol{A};\boldsymbol{X}) = \bar{X}_m \; Sy(n-1;\boldsymbol{B};\boldsymbol{X}/X_m) \vee X_m \; Sy(n-1;\boldsymbol{C};\boldsymbol{X}/X_m), \; (1 \leq m \leq n), \tag{34}$$

where the two sets $\boldsymbol{B}$ and $\boldsymbol{C}$ are both subsets of the set $\boldsymbol{I}_n = \{0, 1, 2, \ldots, n-1\}$, as can be seen from their following definitions

$$\boldsymbol{B} = \boldsymbol{A} \cap \boldsymbol{I}_n, \tag{35}$$

$$\boldsymbol{D} = \{a_0 - 1, a_1 - 1, \ldots, a_m - 1\}, \tag{36}$$

$$\boldsymbol{C} = \boldsymbol{D} \cap \boldsymbol{I}_n. \tag{37}$$

The definitions of the two sets $\boldsymbol{B}$ and $\boldsymbol{C}$ might be restated as follows

$$\boldsymbol{B} = \boldsymbol{A} \qquad if\ a_m \neq n, \tag{38a}$$
$$\boldsymbol{B} = \boldsymbol{A} - \{n\} \qquad if\ a_m = n, \tag{38b}$$

$$\boldsymbol{C} = \boldsymbol{D} \qquad if\ a_0 \neq 0, \tag{39a}$$
$$\boldsymbol{C} = \boldsymbol{D} - \{-1\} \qquad if\ a_0 = 0. \tag{39b}$$

The expansion (34) can be recursively applied till one of the following boundary conditions is reached

$$Sy(n;\boldsymbol{I}_{n+1};\boldsymbol{X}) = \; 1, \tag{40}$$

$$Sy(n;\; \boldsymbol{\phi}\;;\boldsymbol{X}) = \; 0, \tag{41}$$

where $\boldsymbol{I}_{n+1}$ is the universe of discourse (universal set) for $n$ variables, and $\boldsymbol{\phi} = \{\}$ is the empty set (null set or set with no elements).

The two terms in the RHS of (34) is disjoint since $\bar{X}_m$ appears in the first term while $X_m$ appears in the second. Therefore, it is legitimate to replace the OR operator ($\vee$) by an XOR operator ($\oplus$) in (34), namely

$$Sy(n;\boldsymbol{A};\boldsymbol{X}) = \bar{X}_m \; Sy(n-1;\boldsymbol{B};\boldsymbol{X}/X_m) \oplus X_m \; Sy(n-1;\boldsymbol{C};\boldsymbol{X}/X_m), \; (1 \leq m \leq n), \tag{42}$$

Hence, the Boolean derivative of the SSF $Sy(n;\boldsymbol{A};\boldsymbol{X})$ w.r.t. $X_m$ is readily obtained as another SSF given by

$$\frac{\partial Sy(n;A;X)}{\partial X_m} = Sy(n-1;\boldsymbol{B};\boldsymbol{X}/X_m) \oplus Sy(n-1;\boldsymbol{C};\boldsymbol{X}/X_m)\,, \quad (1 \leq m \leq n), \tag{43}$$

$$\frac{\partial Sy(n;A;X)}{\partial X_m} = = Sy(n-1;\boldsymbol{B} \oplus \boldsymbol{C};\boldsymbol{X}/X_m), \quad (1 \leq m \leq n). \tag{44}$$

The total Banzhaf power is given (according to (1), (25) and (44)) by

$$TBP(X_m) = wt\left(\frac{\partial Sy(n;\boldsymbol{A};\boldsymbol{X})}{\partial X_m}\right) = \sum_{a\in \boldsymbol{B}\oplus \boldsymbol{C}} \; c(n-1,a), \quad (1 \le m \le n), \tag{45}$$

where the function $c(n,k)$ denotes the binomial (combinatorial) coefficient, labelled as $n\ choose\ k$ , or the number of ways of choosing $k$ out of $n$ objects when repetition is not allowed and order does not matter. Alternatively, we might compute $TBP(X_m)$ via (5), noting that for a monotonically non-decreasing function $\boldsymbol{B} \sqsubseteq \boldsymbol{C}$, and hence obtain

$$TBP(X_m) = 2\, wt\, (X_m\, (Sy(n;\boldsymbol{A};\boldsymbol{X})/X_m)) - wt\, (Sy(n;\boldsymbol{A};\boldsymbol{X})) = 2\, wt\, (Sy(n-1;\boldsymbol{C};\boldsymbol{X}/X_m)) - (wt(Sy(n-1;\boldsymbol{B};\boldsymbol{X}/X_m) + wt(Sy(n-1;\boldsymbol{C};\boldsymbol{X}/X_m))$$

$$= wt(Sy(n-1;\boldsymbol{C};\boldsymbol{X}/X_m) - wt(Sy(n-1;\boldsymbol{B};\boldsymbol{X}/X_m). \quad (1 \le m \le n). \tag{46}$$

We note, in particular, that we have expressed the weight of the Boolean quotient $(Sy(n;\boldsymbol{A};\boldsymbol{X})/X_m)$ as

$$wt((Sy(n;\boldsymbol{A};\boldsymbol{X})/X_m)) = wt(Sy(n-1;\boldsymbol{C};\boldsymbol{X}/X_m). \tag{47}$$

The normalized total Banzhaf power is given by

$$NTBP(X_m) = \frac{1}{n}, \quad (1 \le m \le n), \tag{48}$$

as expected. In retrospect, we note that we might not have really needed to carry out the aforementioned detailed calculations, because we could have deduced directly from the symmetry of the voting system that the power of the voters is going to be equal Though the computations of this sub-section are not particularly useful for their own sake, they are potentially of notable benefit in handling certain voting systems that possess dominant partial symmetries among voters, which is the case for each of the four voting systems in the next few sections.

We note that under the assumption of a monotone system, the function $Sy(n;\boldsymbol{A};\boldsymbol{X})$ is monotonically non-decreasing, and hence it represents the success $S(k,n,\boldsymbol{X}) = Sy(n;k..n;\boldsymbol{X})$ of a k-out-of-n: G system [64, 80, 81]. It is a symmetric coherent threshold function with unit weights and a threshold equal to $k$, since

$$\{S(k,n,\boldsymbol{X}) = 1\} \quad iff \quad \{\textstyle\sum_{i=1}^{n} X_i \ge k\}. \tag{49}$$

The celebrated majority voting corresponds to $k$ being equal to the ceiling of $(n+1)/2$. A value of $k$ lower than this should naturally not be permitted (practically due to the requirement that $k$ not allowed to be less than half the sum $n$ of weights). However, we might explore the two extreme cases of $k$ that are studied in the reliability literature, namely

(a) The case $k = n$, which represents a series system (no redundancy). This case is barely allowed, since the quota is exactly equal to the sum of weights. There is a single winning coalition (the grand coalition), and every voter possesses veto power. Decisions are taken only unanimously. Strange as it may seems, the series voting system was adopted by the League of Arab States (LAS) (commonly called Arab League) upon its establishment in 1945. Now, majority voting is adopted in the Arab League, with decisions binding only to states who accept them, i.e., the Arab League adopts a peculiar kind of majority rule which fails to bind a state that is outvoted. This might explain why the Arab League has been, so far, shadowy, ineffective, and intrinsically too weak to achieve a significant part of her stated ambitions [82-84]. Some international organizations other than the Arab League, like the World Trade Organization and the Council of the International Seabed Authority also act (at least on some matters) by unanimity rule or consensus [85].

(b) The case $k = 1$, which represents a parallel system (total redundancy). This case is practically avoided, since the quota does not satisfy the requirement of being strictly greater than half the sum of weights. There are $(2^n - 1)$ winning coalitions (including all coalitions except the empty coalition). Under the assumption that voters are independent, this scheme is a ridiculous and haphazard one, in which every voter manages to pass any draft resolution upon his individual wish, only possibly to get it revoked (by some other voter) immediately. This means that a voter might be swinging instantly from a dictator to a dummy, and so on. The fact that this scheme does not work at all is usually presented by proponents of monotheism as a strong argument against polytheism.

### 3. A Small Vector-Weighted System

Consider a voting system adapted from an exercise in Taylor and Pacelli [2] that relates to a family consisting of parents $P_1$ and $P_2$, and three children $X_1$, $X_2$, and $X_3$. When making decisions like whether to go for vacation to the same recreational resort area of last year, or to try somewhere else, they use the following voting system: To pass, any decision must have the support of at least one parent and two children. This voting system is described by a decision switching function, whose minimal or complete sum is

$$f(\boldsymbol{P};\boldsymbol{X}) = f(P_1, P_2; X_1, X_2, X_3) = P_1X_1X_2 \ \vee \ P_1X_2X_3 \ \vee \ P_1X_1X_3 \ \vee \ P_2X_1X_2 \ \vee \ P_2X_2X_3 \ \vee \ P_2X_1X_3. \quad (50)$$

This function has six prime implicants, which correspond to six minimal winning coalitions, namely, $P_1X_1X_2$, $P_1 X_2 X_3$, $P_1X_1X_3$, $P_2X_1X_2$, $P_2 X_2X_3$, and $P_2 X_1X_3$. We observe that the voting system considered is not a scalar-weighted voting system but a vector-weighted one, which means that no one will *ever* find weights and a quota that describe the system. This observation stems from the fact that the system is not swap robust. To see why, we consider the two winning coalitions $P_1X_1X_2$ and $P_2 X_2X_3$, and then swap the voter $P_1$ of the first coalition with the voter $X_3$ of the second coalition. The resulting coalitions are $X_1X_2 X_3$ and $P_1 P_2 X_2$, which are both losing. The first coalition does not have any parent though it has surplus children, while the second coalition has less than two children albeit surplus parents. The vector-weighted nature of the function $f(\boldsymbol{P};\boldsymbol{X})$ is manifested in the fact that it is equivalent not to the assertion of a single inequality, but to the assertion of the conjunction of two separate inequalities, namely

$$\{P_1 + P_2 \geq 1\} \wedge \{X_1 + X_2 + X_3 \geq 1\}. \quad (51)$$

The analysis above demonstrates that generally a bicameral voting system is not a scalar-weighted system but a vector-weighted one. To facilitate further processing, we rewrite the decision function for this voting system in a factored form or a conjunction of two sums of products (allowed by the vector-weighted nature of the system), then replace each of these two sums into a disjoint one, and finally replace the OR operator ($\vee$) by an XOR operator ($\oplus$) in each of these disjoint sums. Mathematically, we obtain

$$f(\boldsymbol{P};\boldsymbol{X}) = (P_1 \ \vee \ P_2)\ (X_1X_2 \ \vee \ X_2X_3 \ \vee \ X_1X_3)$$

$$= (P_1 \vee \overline{P}_1 P_2)\ (X_1X_2 \vee \overline{X}_1X_2X_3 \vee X_1\overline{X}_2X_3)$$

$$= (P_1 \oplus \overline{P}_1 P_2)\ (X_1X_2 \oplus \overline{X}_1X_2X_3 \oplus X_1\overline{X}_2X_3). \quad (52)$$

Due to partial symmetries, we note that $TBP(P_1) = TBP(P_2)$, and $TBP(X_1) = TBP(X_2) = TBP(X_3)$. Hence, it suffices to compute the Boolean derivative w.r.t. one of the two variables $P_1$ and $P_2$ (say $P_1$), and one of the three variables $X_1, X_2$ and $X_3$ (say $X_1$). According to formula (2), we obtain

$$\frac{\partial f}{\partial P_1} = (1 \oplus P_2)\ (X_1X_2 \oplus \overline{X}_1X_2X_3 \oplus X_1\overline{X}_2X_3) = (\overline{P}_2)\ (X_1X_2 \oplus \overline{X}_1X_2X_3 \oplus X_1\overline{X}_2X_3). \quad (53)$$

$$\frac{\partial f}{\partial X_1} = (P_1 \oplus \overline{P}_1 P_2)\ (X_2 \oplus X_2X_3 \oplus \overline{X}_2X_3) = (P_1 \oplus \overline{P}_1 P_2)\ (X_2\overline{X}_3 \oplus \overline{X}_2X_3). \quad (54)$$

which correspond to the total Banzhaf powers of

$$TBP(P_1) = wt\left(\frac{\partial f}{\partial P_1}\right) = (1)(2+1+1) = 4. \quad (55)$$

$$TBP(X_1) = wt\left(\frac{\partial f}{\partial X_1}\right) = (2+1)(1+1) = 6. \quad (56)$$

Alternatively, we could obtain the same results according to formula (5), namely

$$wt(f) = (2+1)(2+1+1) = 12. \quad (57)$$

$$wt((f/P_1)P_1) = wt((P_1)(X_1X_2 \oplus \overline{X}_1X_2X_3 \oplus X_1\overline{X}_2X_3)) = (2)(2+1+1) = 8. \quad (58)$$

$$wt((f/X_1)X_1) = wt((P_1 \oplus \overline{P}_1 P_2)(X_2 \oplus \overline{X}_2X_3)\ X_1) = (2+1)(1+2)(1) = 9. \quad (59)$$

$$TBP(P_1) = (2)(8) - 12 = 4. \quad (60)$$

$$TBP(X_1) = (2)(9) - 12 = 6, \quad (61)$$

or according to formula (6), namely

$$wt((f/\overline{P}_1)\overline{P}_1) = wt((\overline{P}_1P_2)(X_1X_2 \oplus \overline{X}_1X_2X_3 \oplus X_1\overline{X}_2X_3)) = (1)(2+1+1) = 4. \quad (62)$$

$$wt((f/\overline{X}_1)\overline{X}_1) = wt((P_1 \oplus \overline{P}_1 P_2)(\overline{X}_1X_2X_3)) = (2+1)(1) = 3. \quad (63)$$

$$TBP(P_1) = 12 - (2)(4) = 4. \quad (64)$$

$$TBP(X_1) = 12 - (2)(3) = 6, \quad (65)$$

Finally, the vectors of total Banzhaf powers and normalized total Banzhaf powers are

$$\boldsymbol{TBP} = [\,4 \quad 4 \quad 6 \quad 6 \quad 6\,]^T,\ \boldsymbol{NTBP} = \left[\frac{2}{13} \quad \frac{2}{13} \quad \frac{3}{13} \quad \frac{3}{13} \quad \frac{3}{13}\right]^T. \quad (66)$$

It might seem strange that the voting power assigned to a parent is less than that given to a child (a manifestation of parent kindness and graciousness). Actually, the family ‘parliament’ considered is bicameral, with the chamber of parents being easier to approve a decision (requiring just one half of the members). By contrast, the chamber of children requires the relatively harder majority of two thirds to grant its approval. In Section 6, we will view the parents chamber as a 1-out-of-2 system and the children chamber as a 2-out-of-3 system, and we will see that members of the latter system are more powerful since they belong to a system of a tighter region of useful redundancy.

## 4. The United Nations Security Council voting system

The United Nations Security Council (UNSC) is the most powerful UN organ. It adopts a draft resolution on a non-procedural matter if it has the affirmative vote of nine members, including the concurring votes of the five permanent members. This means that a draft does not pass if either it fails to secure nine supporting votes, or a permanent member blocks it by casting a negative vote (exercising veto power). For the sake of simplicity, we ignore the case of abstention or absence of some members.

The initial description of the UNSC voting system did involve weights (a weight of one for each of the fifteen members) and a quota (nine), and it also involved the statement that each of the five permanent members has veto power. Mathematically, this amounts to a conjunction of two inequalities:

$$\{ P_1 \ P_2 \ P_3 \ P_4 \ P_5 \geq 1\} \wedge \{\textstyle\sum_{i=1}^{5} P_i + \sum_{j=1}^{10} N_j \geq 9\}. \qquad (67a)$$

or, equivalently

$$\{ P_1 \ P_2 \ P_3 \ P_4 \ P_5 = 1\} \wedge \{ \textstyle\sum_{j=1}^{10} N_j \geq 4\}. \qquad (67b)$$

The above formulation with two constraints indicates that the dimension of the UNSC voting system is at most 2. It might come as a surprise that this dimension is just 1, as the UNSC voting system can be modeled as a simple scalar-weighted voting system, with no mention whatsoever of veto power, albeit with different quota and weights, including unreasonably excessive weights for members with veto power. The case of the UNSC voting system is a notable demonstration that a voting system in which some members can veto a draft resolution can always be modeled as a scalar-weighted voting system [2]. The UNSC voting system is equivalent (with no members absent or abstaining) to the scalar weighted voting system [39: 7, 7, 7, 7, 7, 1, 1, 1, 1, 1, 1, 1, 1, 1, 1], governed by the inequality

$$\textstyle\sum_{i=1}^{5} 7P_i + \sum_{j=1}^{10} N_j \geq 39. \qquad (68)$$

Note that formula (68) is exactly equivalent to formula (67), though in formula (68) we omitted any mention of veto power, increased the weight of each permanent member seven folds, and correspondingly augmented the quota by $(5)(7-1) = 30$, thereby increasing it from 9 to 39. A winning coalition in the UNSC must contain all the five permanent members (securing a total weight of 35) and at least four non-permanent members (adding a weight of at least 4), thereby attaining a quota of at least 39. A coalition missing a single permanent member and containing all non-permanent ones has a weight of $(4)(7) + (10)(1) = 38$, and hence is losing. A coalition containing all permanent members and three non-permanent ones has a weight of $(5)(7) + (3)(1) = 38$ again, and hence is also losing. Note that the actual UNSC quota is not 9 (out of a total weight of 15) as in formula (67a), but it is 39 (out of a total weight of 45) as in formula (68), which is very high, indeed [85]. The decision function for the UNSC voting system (with no members absent or abstaining) is

$$f(\boldsymbol{P};\boldsymbol{N}) = Sy(5;\{5\},\boldsymbol{P})\ Sy(10;\{4..10\},\boldsymbol{N}) = P_1\ P_2\ P_3\ P_4\ P_5\ Sy(10;\{4..10\},\boldsymbol{N})$$
$$= P_1\ P_2\ P_3\ P_4\ P_5\ (\overline{N}_j\ Sy(9;\{4..9\},\boldsymbol{N}/N_j) \vee N_j\ Sy(9;\{3..9\},\boldsymbol{N}/N_j)) =$$
$$P_1\ P_2\ P_3\ P_4\ P_5\ (\overline{N}_j\ Sy(9;\{4..9\},\boldsymbol{N}/N_j) \oplus N_j\ Sy(9;\{3..9\},\boldsymbol{N}/N_j)). \quad (69)$$

Due to partial symmetries, we note that $TBP(P_i)$ is the same for $(1 \leq i \leq 5)$ and $TBP(N_j)$ is the same for $(1 \leq j \leq 10)$. Hence, it suffices to compute the Boolean derivative w.r.t. one of the five variables $P_i$ (say $P_1$), and one of the ten variables $N_j$ (say $N_1$). According to formula (2), we compute

$$\frac{\partial f}{\partial P_1} = P_2\ P_3\ P_4\ P_5\ Sy(10;\{4..10\};\boldsymbol{N}). \quad (70)$$
$$\frac{\partial f}{\partial N_1} = P_1\ P_2\ P_3\ P_4\ P_5\ (Sy(9;\{4..9\};\boldsymbol{N}/N_1) \oplus Sy(9;\{3..9\};\boldsymbol{N}/N_1))$$
$$= P_1\ P_2\ P_3\ P_4\ P_5\ (Sy(9;\{3\};\boldsymbol{N}/N_1), \quad (71)$$

which correspond to the total Banzhaf powers of (See Appendix C)

$$TBP(P_1) = (1)\sum_{a\in\{4..10\}} c(10,a) = C(10,4) = 1024 - (1 + 10 + 45 + 120) = 848. \quad (72)$$

$$TBP(N_1) = (1)\ c(9,3) = 84. \quad (73)$$

Note that the weight of the 4-variable function ( $P_2\ P_3\ P_4\ P_5$) in (72) is 1, just the same as the weight of the 5-variable function ($P_1\ P_2\ P_3\ P_4\ P_5$) in (73). The total powers of the fifteen members is $(5)(848) + (10)(84) = 5080$. The normalized total Banzhaf powers of a permanent member and a non-permanent one are, respectively, $NTBP(P_i) = 848/5080 = 0.167, (1 \leq i \leq 5)$, and $NTBP(N_j) = 84/5080 = .0165, (1 \leq j \leq 10)$.

We recall that a dummy voter is a voter $V$ who has no say in the outcome of the voting system, since $TBP(V)$ is strictly equal to 0. This voter has no power whatsoever, since he or she cannot influence the passing of a resolution in any case. The existence of a dummy voter defeats the purpose of the voting system, which should allow each individual voter some plausible chance, however small, to affect or influence the decisions made by the system. The UNSC voting system just falls short of assigning dummy status to the non-permanent members, albeit it makes them almost dummies. The $TBP$ of a non-permanent member is alarmingly negligible compared to that of a permanent member but it is not strictly equal to 0. The $TBP$ of all ten non-permanent members of the UNSC put together is slightly less than that of a single permanent member. The voting power in the UNSC is essentially divided into six shares, with five of them divided evenly among the permanent members, and with the sixth share split into sub-shares circulating among alternating representatives of the rest of the world. Napel and Widgrén [86] consider the non-permanent members a special case of what they label as 'inferior players' in the voting game, and hence only permanent members can be viewed as powerful. These authors propose a new voting index called the Strict Power Index (SPI), such that $SPI(X_i) = TBP(X_i)$ for powerful members, and $SPI(X_i) = 0$ for inferior ones, a clever way to strengthen the dummy axiom as the inferior one.

Due to the veto powers of the five permanent members of the present UNSC, the UNSC voting system is severely deficient from all of the aspects of efficiency, equity, effectiveness, legitimacy, and functionality [87-90]. A prominent area of current research deals with the best methodology to reform the UNSC, and the identification of actors or structures that mainly prevent such a reform from materializing [91-97]. Unfortunately, many 'reform' proposals call for a wider membership in the group of permanent and blocking members. Such proposals do not necessarily guarantee the improvement of an already flawed system, but tend to make it more flawed. One proposal that has received some attention is weighted voting [85], but the problem is how to set the weights in a fair way that might be acceptable to all nations.

## 5. A Six-Member Voting System

This section considers the six-member voting system [65; 47, 46, 17, 16, 2, 1], which represents the Scottish Parliament of 2007 [98]. The first five members of this system are political parties, while the sixth is an independent individual. This individual is somehow ignored in many studies (e.g., [99]) that view the reduced five-member system [65; 47, 46, 17, 16, 2]. We first consider the reduced system and then look at the full system. For the reduced system, the decision function might be given by its minimal/complete sum

$$f(\boldsymbol{X}) = X_1X_2 \ \vee \ X_1X_3X_4 \ \vee \ X_1X_3X_5 \ \vee \ X_1X_4X_5 \ \vee \ X_2X_3X_4 \ \vee \ X_2X_3X_5. \quad (74)$$

This function has six prime implicants, which correspond to six minimal winning coalitions ($MWCs$), namely, $X_1X_2$, $X_1X_3X_4$, $X_1X_3X_5$, $X_1X_4X_5$, $X_2X_3X_4$, and $X_2X_3X_5$, each of which being a product of un-complemented literals. We now apply disjointing techniques [64, 67, 71, 74, 77-79] to obtain a disjoint sum for $f(\boldsymbol{X})$, in which the OR operator can legitimately be replaced by the XOR operator, namely

$$f(\boldsymbol{X}) = X_1X_2 \ \oplus \ X_1\overline{X}_2X_3X_4 \ \oplus \ X_1\overline{X}_2X_3\overline{X}_4X_5 \ \oplus \ X_1\overline{X}_2\overline{X}_3X_4X_5 \ \oplus \ \overline{X}_1X_2X_3X_4 \ \oplus \ \overline{X}_1X_2X_3\overline{X}_4X_5. \quad (75)$$

Now, we compute

$$wt(f) = 8+2+1+1+2+1 = 15. \quad (76a)$$

$$wt((f/\overline{X}_1)\overline{X}_1) = wt(X_2X_3X_4 \ \oplus \ X_2X_3\overline{X}_4X_5) = 2+1 = 3. \quad (76b)$$

$$wt\left((f/\overline{X}_2)\overline{X}_2\right) = wt\left(X_1X_3X_4 \ \oplus \ X_1X_3\overline{X}_4X_5 \ \oplus \ X_1\overline{X}_3X_4X_5\right) = 2+1+1 = 4. \quad (76c)$$

$$wt((f/\overline{X}_3)\overline{X}_3) = wt(X_1X_2 \ \oplus \ X_1\overline{X}_2X_4X_5) = 4+1 = 5. \quad (76d)$$

$$wt\left((f/\overline{X}_4)\overline{X}_4\right) = wt\left(X_1X_2 \ \oplus \ X_1\overline{X}_2X_3X_5 \ \oplus \ \overline{X}_1X_2X_3X_5\right) = 4+1+1 = 6. \quad (76e)$$

Note that all the Boolean quotients obtained above remained with disjoint terms (See Appendix D), a feature that facilitated computation of the weights. According to formula (5), we obtain

$$TBP(X_1) = 15-(2)(3) = 9. \quad (77a)$$

$$TBP(X_2) = 15-(2)(4) = 7. \quad (77b)$$

$$TBP(X_3) = 15-(2)(5) = 5. \quad (77c)$$

$$TBP(X_5) = TBP(X_4) = 15-(2)(6) = 3. \quad (77e)$$

Finally, the vectors of total Banzhaf powers and normalized total Banzhaf powers are

$$\boldsymbol{TBP} = [\ 9 \quad 7 \quad 5 \quad 3 \quad 3]^T,\ \boldsymbol{NTBP} = \begin{bmatrix} \frac{9}{27} & \frac{7}{27} & \frac{5}{27} & \frac{3}{27} & \frac{3}{27} \end{bmatrix}^T. \quad (78)$$

The complementary function $\overline{f}(\boldsymbol{X})$ has six prime implicants, which correspond to six maximal losing coalitions ($MLCs$), each of which being a product of un-complemented literals, namely

$$\overline{f}(\boldsymbol{X}) = \overline{X}_1\overline{X}_2 \vee \overline{X}_1\overline{X}_3 \vee \overline{X}_1\overline{X}_4\overline{X}_5 \vee \overline{X}_2\overline{X}_3\overline{X}_4 \vee \overline{X}_2\overline{X}_3\overline{X}_5 \vee \overline{X}_2\overline{X}_4\overline{X}_5. \quad (79)$$

Note that a prime implicant of $\overline{f}$ such as $\overline{X}_1\overline{X}_2$ signifies that the first two members are opposing the draft resolution, and hence it corresponds to a maximal losing coalition ($MLC$) that consists of the remaining members $\{X_3, X_4, X_5\}$. This coalition is losing since its total weight $(W_3 + W_4 + W_5) = 17 + 16 + 2 = 35$, which is less the quota of $65$. It is maximal since it swings from being losing to being winning when augmented with one extra voter (either $X_1$ or $X_2$). We now apply disjointing techniques to obtain a disjoint sum for $\overline{f}(\boldsymbol{X})$, in which the OR operator can legitimately be replaced by the XOR operator, namely

$$\overline{f}(\boldsymbol{X}) = \overline{X}_1\overline{X}_2 \oplus \overline{X}_1 X_2\overline{X}_3 \oplus \overline{X}_1X_2X_3\overline{X}_4\overline{X}_5 \oplus X_1\overline{X}_2\overline{X}_3\overline{X}_4 \oplus X_1\overline{X}_2\overline{X}_3X_4\overline{X}_5 \oplus X_1\overline{X}_2X_3\overline{X}_4\overline{X}_5. \quad (80)$$

Now, we repeat the previous computations in terms of the complemented function $\overline{f}(\boldsymbol{X})$ and its Boolean quotients w.r.t. the uncomplemented $X_m,\ 1 \leq m \leq 4$.

$$wt(\overline{f}) = 8 + 4 + 1 + 2 + 1 + 1 = 17. \quad (81a)$$

$$wt\left((\overline{f}/X_1)X_1\right) = wt(\overline{X}_2\overline{X}_3\overline{X}_4 \oplus \overline{X}_2\overline{X}_3X_4\overline{X}_5 \oplus \overline{X}_2X_3\overline{X}_4\overline{X}_5) = 2 + 1 + 1 = 4. \quad (81b)$$

$$wt\left((\overline{f}/X_2)X_2\right) = wt(\overline{X}_1\overline{X}_3 \oplus \overline{X}_1X_3\overline{X}_4\overline{X}_5) = 4 + 1 = 5. \quad (81c)$$

$$wt((\overline{f}/X_3)X_3) = wt(\overline{X}_1\overline{X}_2 \oplus \overline{X}_1X_2\overline{X}_4\overline{X}_5 \oplus X_1\overline{X}_2\overline{X}_4\overline{X}_5) = 4 + 1 + 1 = 6. \quad (81d)$$

$$wt\left((\overline{f}/X_4)X_4\right) = wt(\overline{X}_1\overline{X}_2 \oplus \overline{X}_1X_2\overline{X}_3 \oplus X_1\overline{X}_2\overline{X}_3\overline{X}_5) = 4 + 2 + 1 = 7. \quad (81e)$$

Note that $wt(f)$ and $wt(\overline{f})$ add to $2^5 = 32$, as required, and that all the Boolean quotients obtained above remained with disjoint terms, a feature that facilitated computation of the weights. According to formula complementary to (5), we recover our earlier results

$$TBP(X_1) = 17 - (2)(4) = 9. \quad (82a)$$

$$TBP(X_2) = 17 - (2)(5) = 7. \quad (82b)$$

$$TBP(X_3) = 17 - (2)(6) = 5. \quad (82c)$$

$$TBP(X_5) = TBP(X_4) = 17 - (2)(7) = 3. \quad (82d)$$

The analysis above demonstrates that it is immaterial whether the decision function $f(\boldsymbol{X})$ or its complement $\overline{f}(\boldsymbol{X})$ is used in the computation of the Banzhaf index. Some other voting-power indices do not share this property. An obvious example is the Public Good Index (PGI) due to Holler [47, 48, 99]. It might be worth noting that Manfred Holler (who is definitely one of the most prominent contemporary game theorists) could have been possibly intrigued by the possibility that had it not been for his (presumed) lack of prominence, his Public Good Index (PGI) would have been as popular as the Banzhaf index [99]. We note that the $PGI$ looks at $f(\boldsymbol{X})$ so as to count the number of the products representing the $MWCs$ in which the uncomplemented literal of a certain voter appears, and take this number as the power of this voter. For the present example, $\boldsymbol{PGI} = [\,4 \quad 3 \quad 4 \quad 3 \quad 3]^T$. If instead, it is made to look at $\overline{f}(\boldsymbol{X})$, to count the number of the products representing the $MLCs$ in which the complemented literal of the voter appears, a complementary index (that we denote $CPGI$) emerges. For the present example, $\boldsymbol{CPGI} = [\,3 \quad 4 \quad 3 \quad 3 \quad 3]^T$, which is clearly different. Of course, it is possible to combine the $PGI$ and the $CPGI$ (or normalized versions thereof) by considering their arithmetic, geometric, or harmonic means. Neither $PGI$ nor $CPGI$ keeps local monotonicity with the weights (as the Banzhaf index does). The monotonicity property for a voting index does not allow a voter of a larger weight to enjoy less voting power than a voter with a smaller voting weight.

The decision function $f(\boldsymbol{X})$ of the extended 6-member system [65; 47, 46, 17, 16, 2, 1] is just that of the reduced 5-member system [65; 47, 46, 17, 16, 2], augmented with two extra $MWCs$, namely $X_1X_3X_6$ and $X_2X_4X_5\,X_6$. The expression (74) for $f(\boldsymbol{X})$ should be ORed with $(X_1X_3X_6 \ \vee \ X_2X_4X_5\,X_6)$, while its disjoint expression (75) need be XORed with $(X_1\overline{X}_2X_3\overline{X}_4\overline{X}_5X_6 \ \oplus \ \overline{X}_1X_2\overline{X}_3X_4X_5\,X_6)$. Had $f(\boldsymbol{X})$ retained its expression, its weight would have been simply doubled, because it is now a 6-variable function rather than a 5-variable one. But now we need (after doubling the weight from 15 to 30) to augment its value with $wt\big(X_1\overline{X}_2X_3\overline{X}_4\overline{X}_5X_6 \ \oplus \ \overline{X}_1X_2\overline{X}_3X_4X_5\,X_6\big) = 1 + 1 = 2$ to make it 32. Now, we repeat calculation for the quantities in (76b)-(76e), by doubling their original values and adding the weight of the appropriate Boolean quotient of $\big(X_1\overline{X}_2X_3\overline{X}_4\overline{X}_5X_6 \ \oplus \ \overline{X}_1X_2\overline{X}_3X_4X_5\,X_6\big)$, provided disjointness is retained. Specifically we obtain

$$wt((f/\overline{X}_1)\overline{X}_1) = (2)(3) + wt(\,X_2\overline{X}_3X_4X_5\,X_6) = 6 + 1 = 7. \tag{83a}$$

$$wt\Big((f/\overline{X}_2)\overline{X}_2\Big) = (2)(4) + wt\big(X_1X_3\overline{X}_4\overline{X}_5X_6\big) = 8 + 1 = 9. \tag{83b}$$

$$wt((f/\overline{X}_3)\overline{X}_3) = (2)(5) + wt(\,\overline{X}_1X_2X_4X_5\,X_6) = 10 + 1 = 11. \tag{83c}$$

$$wt\Big((f/\overline{X}_4)\overline{X}_4\Big) = (2)(6) + wt\big(X_1\overline{X}_2X_3\overline{X}_5X_6\big) = 12 + 1 = 13. \tag{83d}$$

Moreover, we develop the new term $wt\Big((f/\overline{X}_6)\overline{X}_6\Big)$ as the weight of the original 5-variable $f$, which is 15. According to formula (5), we obtain

$$TBP(X_1) = 32 - (2)(7) = 18. \tag{84a}$$

$$TBP(X_2) = 32 - (2)(9) = 14. \tag{84b}$$

$$TBP(X_3) = 32 - (2)(11) = 10. \tag{84c}$$

$$TBP(X_5) = TBP(X_4) = 32 - (2)(13) = 6. \tag{84d}$$

$$TBP(X_6) = 32 - (2)(15) = 2. \tag{84e}$$

Finally, the vectors of total Banzhaf powers and normalized total Banzhaf powers are

$$\boldsymbol{TBP} = [\,18 \quad 14 \quad 10 \quad 6 \quad 6 \quad 2\,]^T, \boldsymbol{NTBP} = [\tfrac{18}{56} \quad \tfrac{14}{56} \quad \tfrac{10}{56} \quad \tfrac{6}{56} \quad \tfrac{6}{56} \quad \tfrac{2}{56}]^T. \tag{85}$$

The complementary function $\overline{f}(\boldsymbol{X})$ of the extended 6-member system [65; 47, 46, 17, 16, 2, 1] now has eight prime implicants, namely

$$\overline{f}(\boldsymbol{X}) = \overline{X}_1\overline{X}_2 \vee \overline{X}_1\overline{X}_3\overline{X}_4 \vee \overline{X}_1\overline{X}_3\overline{X}_5 \vee \overline{X}_1\overline{X}_3\overline{X}_6 \vee \overline{X}_1\overline{X}_4\overline{X}_5 \vee \overline{X}_2\overline{X}_3\overline{X}_4 \vee \overline{X}_2\overline{X}_3\overline{X}_5 \vee \overline{X}_2\overline{X}_4\overline{X}_5\overline{X}_6. \tag{86}$$

We now apply disjointing techniques to obtain a disjoint sum for $\overline{f}(\boldsymbol{X})$, in which the OR operator can legitimately be replaced by the XOR operator, namely

$$\overline{f}(\boldsymbol{X}) = \overline{X}_1\overline{X}_2 \oplus \overline{X}_1X_2\overline{X}_3\overline{X}_4 \oplus \overline{X}_1X_2\overline{X}_3X_4\overline{X}_5 \oplus \overline{X}_1X_2\overline{X}_3X_4X_5\overline{X}_6 \oplus \overline{X}_1X_2X_3\overline{X}_4\overline{X}_5 \oplus X_1\overline{X}_2\overline{X}_3\overline{X}_4 \oplus X_1\overline{X}_2\overline{X}_3X_4\overline{X}_5 \oplus X_1\overline{X}_2X_3\overline{X}_4\overline{X}_5\overline{X}_6. \tag{87}$$

Now, we repeat the previous computations in terms of the complemented function $\overline{f}(\boldsymbol{X})$ and its Boolean quotients w.r.t. the uncomplemented $X_m,\ m = 1, 2, 3, 4, 6$ .

$$wt(\overline{f}) = 16 + 4 + 2 + 1 + 2 + 4 + 2 + 1 = 32. \tag{88a}$$

$$wt\left((\overline{f}/X_1)X_1\right) = wt(\overline{X}_2\overline{X}_3\overline{X}_4 \oplus \overline{X}_2\overline{X}_3X_4\overline{X}_5 \oplus \overline{X}_2X_3\overline{X}_4\overline{X}_5\overline{X}_6) = 4 + 2 + 1 = 7. \tag{88b}$$

$$wt\left((\overline{f}/X_2)X_2\right) = wt(\overline{X}_1\overline{X}_3\overline{X}_4 \oplus \overline{X}_1\overline{X}_3X_4\overline{X}_5 \oplus \overline{X}_1\overline{X}_3X_4X_5\overline{X}_6 \oplus \overline{X}_1X_3\overline{X}_4\overline{X}_5) = 4 + 2 + 1 + 2 = 9. \tag{88c}$$

$$wt((\overline{f}/X_3)X_3) = wt(\overline{X}_1\overline{X}_2 \oplus \overline{X}_1X_2\overline{X}_4\overline{X}_5 \oplus X_1\overline{X}_2\overline{X}_4\overline{X}_5\overline{X}_6) = 8 + 2 + 1 = 11. \tag{88d}$$

$$wt\left((\overline{f}/X_4)X_4\right) = wt(\overline{X}_1\overline{X}_2 \oplus \overline{X}_1X_2\overline{X}_3\overline{X}_5 \oplus \overline{X}_1X_2\overline{X}_3X_5\overline{X}_6 \oplus X_1\overline{X}_2\overline{X}_3\overline{X}_5) = 8 + 2 + 1 + 2 = 13. \tag{88e}$$

$$wt\left((\overline{f}/X_6)X_6\right) = wt(\overline{X}_1\overline{X}_2 \oplus \overline{X}_1X_2\overline{X}_3\overline{X}_4 \oplus \overline{X}_1X_2\overline{X}_3X_4\overline{X}_5 \oplus \overline{X}_1X_2X_3\overline{X}_4\overline{X}_5 \oplus X_1\overline{X}_2\overline{X}_3\overline{X}_4 \oplus X_1\overline{X}_2\overline{X}_3X_4\overline{X}_5\,) = 8 + 2 + 1 + 1 + 2 + 1 = 15. \tag{88f}$$

Note that $wt(f)$ and $wt(\overline{f})$ add to $2^6 = 64$, as required, and that all the Boolean quotients obtained above remained with disjoint terms, a feature that facilitated computation of the weights. According to a formula complementary to formula (5), we recover our earlier results

$$TBP(X_1) = 32 - (2)(7) = 18. \tag{89a}$$

$$TBP(X_2) = 32 - (2)(9) = 14. \tag{89b}$$

$$TBP(X_3) = 32 - (2)(11) = 10. \tag{89c}$$

$$TBP(X_5) = TBP(X_4) = 32 - (2)(13) = 6. \tag{89d}$$

$$TBP(X_6) = 32 - (2)(15) = 2. \quad (89e)$$

The analysis above demonstrates again that it is immaterial whether the decision function $f(\boldsymbol{X})$ or its complement $\overline{f}(\boldsymbol{X})$ is used in the computation of the Banzhaf index. For the extended 6-member system [65; 47, 46, 17, 16, 2, 1], the Public Good Index and its complement are given by

$$\boldsymbol{PGI} = [\,5 \quad 4 \quad 5 \quad 4 \quad 4 \quad 2\,]^T, \boldsymbol{CPGI} = [\,5 \quad 4 \quad 5 \quad 4 \quad 4 \quad 2\,]^T. \quad (90)$$

This time the two indices are the same, albeit they still lack local monotonicity with the weights (while the Banzhaf index does).

## 6. A Tri-cameral Parliament of k-out-of-n Chambers

In this section, we consider a fictitious tri-cameral parliament modeled by a vector-weighted voting system of dimension 3. The chambers of this parliament are a $k_1$-out-of-$n_1$ system of voters $\boldsymbol{X}$, a $k_2$-out-of-$n_2$ system of voters $\boldsymbol{Y}$, and a $k_3$-out-of-$n_3$ system of voters $\boldsymbol{Z}$, respectively. Due to symmetry, the voting powers of members of one chamber are the same, and it suffices to compute the power of a single member of each chamber. The minimal sum for the decision function $f(\boldsymbol{X},\boldsymbol{Y},\boldsymbol{Z})$ (which happens to be also its complete sum since it is unate) is

$$f(\boldsymbol{X},\boldsymbol{Y},\boldsymbol{Z}) = Sy(n_1;\{k_1..n_1\};\boldsymbol{X}) \wedge Sy(n_2;\{k_2..n_2\};\boldsymbol{Y}) \wedge Sy(n_3;\{k_3..n_3\};\boldsymbol{Z}), \quad (91)$$

while that for the complement function is

$$\overline{f}(\boldsymbol{X},\boldsymbol{Y},\boldsymbol{Z}) = Sy(n_1;\{0..k_1-1\};\boldsymbol{X}) \vee Sy(n_2;\{0..k_2-1\};\boldsymbol{Y}) \vee Sy(n_3;\{0..k_3-1\};\boldsymbol{Z}). \quad (92)$$

The Boole-Shannon expansion for the constituent SSF $Sy(n_1;\{k_1..n_1\};\boldsymbol{X})$ in (91) about any of its variables, say $X_m$ $(1 \le m \le n_1)$ can be produced from (34) as follows

$$Sy(n_1;\{k_1..n_1\};\boldsymbol{X}) = \overline{X}_m\ Sy(n_1-1;\{k_1..n_1-1\};\boldsymbol{X}/X_m) \vee X_m\ Sy(n_1-1;\{k_1-1..n_1-1\};\boldsymbol{X}/X_m),\ (1 \le m \le n_1), \quad (93)$$

Hence, we can use (25), (93), (6), (C.4), and (C.6) to compute the following

$$wt(f) = \left(\textstyle\sum_{a\in\{k_1..n_1\}} c(n_1,a)\right) \left(\textstyle\sum_{a\in\{k_2..n_2\}} c(n_2,a)\right) \left(\textstyle\sum_{a\in\{k_3..n_3\}} c(n_3,a)\right) = C(n_1,k_1)\ C(n_2,k_2)\ C(n_3,k_3). \quad (94)$$

$$wt(f/\overline{X}_m) = wt(Sy(n_1-1;\{k_1..n_1-1\};\boldsymbol{X}/X_m) \wedge Sy(n_2;\{k_2..n_2\};\boldsymbol{Y}) \wedge Sy(n_3;\{k_3..n_3\};\boldsymbol{Z})) = C(n_1-1,k_1)\,C(n_2,k_2)\,C(n_3,k_3). \quad (95)$$

$$TBP(X_m) = (C(n_1,k_1) - 2\,C(n_1-1,k_1))\ C(n_2,k_2)\ C(n_3,k_3)$$

$$= (C(n_1-1,k_1-1) - C(n_1-1,k_1))\ C(n_2,k_2)\ C(n_3,k_3)$$

$$= c(n_1-1,k_1-1)\ C(n_2,k_2)\ C(n_3,k_3). \quad (96)$$

where $c(n,k)$ is the combinatorial coefficient and $C(n,k) = \sum_{m=k}^{n} c(n,m)$ is the cumulative combinatorial coefficient (See Appendix C). Similarly, we obtain

$$TBP(Y_m) = \ C(n_1, k_1)\ c(n_2 - 1, k_2 - 1)\ C(n_3, k_3), \tag{97}$$

$$TBP(Z_m) = \ C(n_{1,}, k_1)\ C(n_2, k_2)\ c(n_3 - 1, k_3 - 1). \tag{98}$$

The tri-cameral system considered might be reduced to a bi-cameral one if we assume that the third chamber is the always-approving 0-out-of-0 system, with $Sy(0; \{0..0\}; \boldsymbol{Z}) = 1$ and $C(0,0) = 1$. In this case, we obtain

$$TBP(X_m) = \ c(n_1 - 1, k_1 - 1)\ C(n_2, k_2). \tag{99}$$

$$TBP(Y_m) = \ C(n_1, k_1)\ c(n_2 - 1, k_2 - 1). \tag{100}$$

We can now replicate the results of the past two sections. The family parliament of Section 3 is the intersection of a 1-out-of-2 system and a 2-out-of-3 system. Hence

$$TBP(P_1) = c(1,0)\ C(3,2) = (1)(4) = 4. \tag{101}$$

$$TBP(X_1) = C(2,1)\ c(2,1) = (3)(2) = 6, \tag{102}$$

The UNSC of Section 4 is the intersection of a 5-out-of-5 system and a 4-out-of-10 system. Hence

$$TBP(P_1) = c(4,4)\ C(10,4) = (1)(848) = 848, \tag{103}$$

$$TBP(N_1) = C(5,5)\ c(9,3) = (1)(84) = 84. \tag{104}$$

Now, let us consider a tri-cameral system that is the intersection of a 7-out-of-9 system, a 5-out-of-7 system and a 3-out-of-5 system. The Banzhaf indices of sample members of the three chambers are

$$TBP(X_m) = c(8,6)C(7,5)\ C(5,3) = (28)(29)(16) = 12{,}992, \tag{105}$$

$$TBP(Y_m) = C(9,7)c(6,4)\ C(5,3) = (46)(15)(16) = 11{,}040, \tag{106}$$

$$TBP(Z_m) = \ C(9,7)C(7,5)\ c(4,2) = (46)(29)(6) = 8{,}004. \tag{107}$$

We digress a little bit to consider the region of useful redundancy for a k-out-of-n system of identical components, which is the interval in component reliability for which the reliability of the system is better than that of a single component [100-102]. This region is given for the three chambers considered as $(0.842, 1.00)$, $(0.744, 1.00)$, and $(0.500, 1.00)$, respectively [100]. Our calculations suggest that most powerful among members of the present tri-cameral parliament are those of the chamber whose k-out-of-n system possesses the narrowest region of useful redundancy. This means that the weaker the redundancy of the voting system, the stronger the voting power enjoyed by its voters. If one of the three chambers is a series system (no redundancy), then its members are more powerful than these of the other chambers. Dually, if one of the three chambers is a parallel system (total redundancy), then its members are less powerful than these of the other chambers.

We might also observe that only the third chamber of the aforementioned system (the 3-out-of-5 one) employs a simple majority rule. By contrast, the first two chambers impose a supermajority rule, which is a departure from simple majority towards unanimity or consensus. Here, the 7-out-

of-9 chamber demands a higher supermajority than that required by the 5-out-of-7 chamber. As the supermajority required by a chamber increases, the decision costs increase as well (because more voters must agree to pass a draft resolution, and it becomes harder to pass a decision) but the costs of majority-versus-minority exploitation decline (because fewer voters can be outvoted or exploited, and a voter becomes more powerful) [85].

We can use our present results in an admittedly oversimplified analysis of the United States (US) federal voting system, which is a 537-variable system consisting of the President $P$, the Vice-President $V$, $S_i$ or senator $i$ (member $i$ of the Senate) $(1 \leq i \leq 100)$, and $H_j$ or member $j$ of the House of Representatives $(1 \leq j \leq 435)$ [2, 18]. If we ignore the power of the President and that of the Vice-President, the problem reduces to one of a bi-cameral parliament and should be solved by equations (99) and (100). To be realistic, we separate the problem into two cases. In the first case, the President does not exercise his veto power, and the federal system reduces to an intersection of the Senate as a 51-out-of-100 system, and the House as 218-out-of-435 system (i.e., with simple majority in each chamber). For this case, we obtain

$$TBP(S_i) = c(99, 50)\, C(435, 218) = 5.045\text{e}(+28) \times 4.436\text{e}(+130) = 2.238\text{e}(+159), \quad (108)$$

$$TBP(H_j) = C(100, 51)\, c(434, 217) = 5.834e(+29) \times 1.698\text{e}(+129) = 9.906\text{e}(+158). \quad (109)$$

In the second case, the President exercises his veto power, and the federal system reduces to an intersection of the Senate as a 67-out-of-100 system, and the House as 290-out-of-435 system (i.e., with two-thirds majority in each chamber). For this case, we obtain

$$TBP(S_i) = c(99, 66)\, C(435, 290) = 1.974e(+26) \times 1.420e(+119) = 2.804e(+145). \quad (110)$$

$$TBP(H_j) = C(100, 67)\, c(434, 289) = 5.538e(+26) \times 4.797e(+118) = 2.656e(+145). \quad (111)$$

Taylor and Pacelli [2] handled the above problem by neglecting the power of the Vice-President only. They obtained $NTBP(P) = 0.038$, $NTBP(S_i) = 0.0033$, and $NTBP(H_j) = 0.0015$. According to [2], the ratio of the power of a senator to that of a representative $(TBP(S_i)/TBP(H_j))$ is $2.200$, while our approximate calculations set this ratio at $2.259$ and $1.056$, depending on whether the president does not or does cast a veto.

## 7. Conclusions

This paper is a major step within an ongoing project [8, 10, 11, 16-18] that strives to construct a switching-algebraic theory for weighted monotone voting systems. This new theory is naturally expected to be a complement rather than a replacement of the more dominant and all-encompassing game-theoretic description of these systems. Though this paper is entirely devoted to the exploration of a prominent index of voting power, *viz.*, the Banzhaf Index, we had a chance to add a few remarks about two other (albeit related) indices, namely: (a) the Strict Power Index (SPI) of Napel and Widgrén [86], and (b) the Public Good Index (PGI) of Holler (and Packel) [47,48, 99]. The paper serves as a tutorial exposition of the computation of pertinent switching-algebraic quantities, including Boolean differencing (Boolean differentiation), Boolean quotient construction (Boolean restriction), and computation of the weight (the number of true vectors or minterms) of a switching function. The paper also tries to make the most of symmetry features through competent utilization of the properties and expansions of symmetric switching functions.

Detailed (and multiple) solutions are offered for four examples, which covered both scalar-weighted and vector-weighted voting systems.

Throughout this paper, we experimented with several formulas for obtaining the Banzhaf indices for a monotone system. Among these formulas, equation (2) utilizes partial differentiation of a switching function, while equations (5), (6), and (7) rely on the construction of Boolean quotients. All afore-mentioned equations require the computation of weights of switching functions, a task that is facilitated by casting these functions in disjoint sum-of-products form. Equations (5)-(7) enjoy the advantage that their construction of Boolean quotients preserves disjointness, and hence facilitates weight computation. This advantage is missing for equation (2), in which partial differentiation occasionally destroys disjointness.

In the analysis employed in the present paper, we made the explicit assumption that variables in any considered system are statistically independent. For future work, we need to relax this assumption, and to consider (a) the issue of partisan identification and commitment [103-105], which leads to similar voting patterns among many voters (analogous to common-cause effect in reliability studies), and (b) the issue of incompatibility among certain voters [106-108], which nullifies any potential coalition that would have included these voters simultaneously. Another prospective area is to integrate the switching-algebraic aspects of voting theory and Qualitative Comparative Analysis (QCA) [14, 52, 109]. Haake and Schneider [110] set the stage for this integration by suggesting that the Banzhaf Index measures the explanatory power of a single condition, averaged across all sufficient combinations of conditions, and hence proposing that this index can serve as an additional goodness-of-fit parameter in QCA. Moreover, since minimal winning coalitions ($MWCs$) are directly represented by switching-algebraic expressions, it seems opportune to investigate earlier studies that relate $MWCs$ to voting power [8, 23, 24, 47, 48, 111-114].

Voting systems translate the idea that voters are not all equal by assigning them different weights. In such a situation, two voters are symmetric in a game if interchanging the two voters leaves the outcome of the game unchanged. Two voters with the same weight are naturally symmetric in every weighted voting game, but the converse statement is not necessarily true. In fact, the power/influence distribution is not a faithful replica of the weight distribution. The two distributions are not necessarily in exact agreement, though they are in close or good agreement for most of the voting systems (according to most power indices) [115]. Such a scenario of inconsistent weighting is the reason behind the ‘design problem’ or ‘inverse problem’ in the power index literature. This problem might be formulated as follows: “Given a target vector of voters’ power or influence and a specific power index, determine a scalar-weighted voting game such that the distribution of power/influence among the various voters is *as close as possible* to the given target vector [116-120].” In this formulation, power indices are utilized in a normative perspective rather than from a descriptive one, and we opt for the description ‘as close as possible’ rather than ‘exactly.’ In fact, we cannot always find a weight assignment that accurately reflects desired power/influence [121]. However, there are certain weighted voting systems, for which some choice of voting weights accurately reflect pre-defined power/influence [7, 121].

In a different line of future work, we plan to employ switching-algebraic techniques to tackle other standard voting systems. A prominent and computationally-intensive system to be considered is the vector-weighted 537-variable system that describes the federal voting system of the United States of America [2, 18]. We offered an approximate solution herein for the ratio of the power of a senator to that of a representative, when the roles of the President and Vice-President are ignored. However, we need to obtain an exact solution that finds the power of everyone involved, especially the President. Another prominent example to consider is the voting system of the Council of the European Union in the period 2014–2020 under the treaty of Lisbon [31-34, 122]. This system requires the *disjunction* (rather than the conjunction) of two conditions (a) the support of at least 16 member states that represent at least 65% of the total EU population, or (b) the support of at least 25 out of the (then) 28 member states [122]. We are going also to explore some of the paradoxes associated with voting powers, such as the paradox of redistribution, the paradox of new members, the quarrelling paradox, the donation paradox, and the paradox of large size [123-129]. We believe that after completing the aforementioned tasks, we will be in a position to tell the entire tale of voting systems from a purely switching-algebraic perspective.

## Appendix A: Unate Boolean Functions

A Boolean function $f(\boldsymbol{X}) = f(X_1, X_2, \ldots, X_{m-1}, X_m, X_{m+1}, \ldots, X_n)$ is called *unate* if and only if (iff) it can be represented as a normal (sum-of-products or product-of-sums) formula in which no variable appears both complemented and un-complemented, *i.e.*, such that every variable is mono-form and no variable is bi-form. This means that a sum-of-products (a product-of-sums) formula for a unate function has zero opposition between any two of its terms or products (alterms or sums). A unate Boolean function is called *positive* in its argument $X_m$, if there exists a normal representation of the function in which $X_m$ does not appear complemented. This happens if and only if every occurrence of the literal $\bar{X}_m$ is redundant and can be eliminated, *i.e.*, if and only if there exist functions $f_{m1}$ and $f_{m2}$ (independent of $X_m$) such that [14-18]

$$f(\boldsymbol{X}) = X_m \ f_{m1}(X_1, X_2, \ldots, X_{m-1}, X_{m+1}, \ldots, X_n) \vee \ f_{m2}(X_1, X_2, \ldots, X_{m-1}, X_{m+1}, \ldots, X_n). \quad \text{(A.1)}$$

denoted as $(f = X_m \ f_{m1} \vee \ f_{m2})$, for short. A unate Boolean function that has positive polarity for each of its arguments $X_m$ is noted to be monotonically non-decreasing (monotonically increasing, for short), and is used to represent the decision of a monotone scalar-weighted voting system [8] or the success of a coherent threshold reliability system [6, 7, 15].

If the function $f(\boldsymbol{X})$ is *positive* in its argument $X_m$, then its subfunctions are $f(\boldsymbol{X}|1_\mathrm{m}) = f(\boldsymbol{X})/X_m = f_{m1} \vee f_{m2}$ and $f(\boldsymbol{X}|0_\mathrm{m}) = f(\boldsymbol{X})/\bar{X}_m = f_{m2}$, which means that $f(\boldsymbol{X}|0_\mathrm{m}) \leq f(\boldsymbol{X}|1_\mathrm{m})$. The subfunctions in the Boole-Shannon expansion of a unate function are also unate functions possessing the same type of polarity for the remaining arguments (positive, negative, or mixed). All threshold (linearly-separable) functions are unate, but the converse is not true. The function $X_1 X_2 \vee X_3 X_4$ is an example of a unate function that is not threshold. All the prime implicants of a unate function are essential, so that it has a single irredundant disjunctive form, which serves as both its (unique) minimal sum and its complete sum. The converse is not necessarily true.

In passing, we stress that we restrict our analysis herein to monotone voting systems, whose decision functions $f(\boldsymbol{X})$ are positively-polarized unate functions. In our mathematical processing of these functions (e.g., expansion, differentiation, and disjointing) we might create formulas for these functions that contain complemented literals of some of its variables. However, the function retains its unateness property. By contrast, if we subject the function $f(\boldsymbol{X})$ to certain restrictions, we might spoil its unateness (at least partially). For example, if we mandate that voters $X_m$ and $X_k$ never be in the same coalition, we need to nullify any instance of the product $X_m\, X_k$ in $f(\boldsymbol{X})$. Such nullification destroys the unateness of $f(\boldsymbol{X})$ w.r.t. each of the two variables $X_m$ and $X_k$, but does not spoil its mono-polarization in the rest of the variables.

## Appendix B: Proof for the Alternative Formulas of the Banzhaf Index for Monotonically Non-Decreasing Boolean Functions

The weight of a monotonically non-decreasing Boolean function is obtained via (A.1) and (20) as

$$
\begin{aligned}
wt(f) &= wt\,(X_m\ f_{m1} \vee\ f_{m2}) = wt\,(X_m\,(f_{m1} \vee\ f_{m2}) \oplus \bar{X}_m\ f_{m2}) \\
&= wt\,(X_m\,(f_{m1} \vee f_{m2})) + wt(\bar{X}_m\, f_{m2}) = wt\,(X_m)\, wt\,(f_{m1} \vee f_{m2}) + wt(\bar{X}_m)\, wt(f_{m2}) \\
&= (1)\, wt\,(f_{m1} \vee f_{m2}) + (1)\, wt(f_{m2}) = wt\,(f_{m1} \vee f_{m2}) + wt(f_{m2}) \\
&= wt(f_{m1}) + wt(f_{m2}) - wt(f_{m1}\, f_{m2}) + wt(f_{m2}) \\
&= wt(f_{m1}) + 2\, wt(f_{m2}) - wt(f_{m1}\, f_{m2}).
\end{aligned} \tag{B.1}
$$

The partial derivative of $f(\boldsymbol{X}) = (X_m\,(f_{m1} \vee\ f_{m2}) \oplus\ \bar{X}_m\ f_{m2})$ w.r.t. $X_m$ is

$$
(f_{m1} \vee\ f_{m2}) \oplus\ f_{m2} = (f_{m1} \vee\ f_{m2})\,\overline{f_{m2}}\ \vee\ f_{m2}\,\overline{f_{m1}}\ \overline{f_{m2}} = f_{m1}\overline{f_{m2}}. \tag{B.2}
$$

The total Banzhaf power of voter number $m$ $(1 \le m \le n)$ according to equation (2) is

$$
TBP(X_m) = wt\left(\frac{\partial f(\boldsymbol{X})}{\partial X_m}\right) = wt\big(f_{m1}\overline{f_{m2}}\big) = wt(f_{m1}) - wt(f_{m1}\, f_{m2}). \tag{B.3}
$$

The first term in equation (5) involves the weight

$$
wt((f(\boldsymbol{X})/X_m)X_m) = wt\,(X_m\ (f_{m1} \vee f_{m2})) = wt\,(X_m)\, wt\ (f_{m1} \vee\ f_{m2}) = wt(f_{m1}) + wt(f_{m2}) - wt(f_{m1}\, f_{m2}).
$$

Therefore, the alternative expression for $TBP(X_m)$ in (5) is given for $(1 \le m \le n)$ by

$$
\begin{aligned}
TBP(X_m) &= 2\, wt((f(\boldsymbol{X})/X_m)X_m) - wt\big(f(\boldsymbol{X})\big) = 2(wt(f_{m1}) + wt(f_{m2}) - wt(f_{m1}\, f_{m2})) - \\
&\big(wt(f_{m1}) + 2\, wt(f_{m2}) - wt(f_{m1}\, f_{m2})\big) = wt(f_{m1}) - wt(f_{m1}\, f_{m2}).
\end{aligned} \tag{B.4}
$$

Equations (B.3) and (B.4) exactly agree in their expression of the Banzhaf index $TBP(X_m)$ for $(1 \le m \le n)$. Now, the second term in equation (6) involves the weight

$$
wt(\big(f(\boldsymbol{X})/\overline{X}_m\big)\overline{X}_m) = wt(\bar{X}_m\ f_{m2}) = wt(f_{m2}). \tag{B.5}
$$

Therefore, the alternative expression for $TBP(X_m)$ in (6) is given for $(1 \le m \le n)$ by

$$TBP(X_m) = wt(f(\boldsymbol{X})) - 2\, wt\left((f(\boldsymbol{X})/\overline{X}_m)\overline{X}_m\right) = wt(f_{m1}) + 2\, wt(f_{m2}) - wt(f_{m1}\, f_{m2}) - 2wt(f_{m2}) = wt(f_{m1}) - wt(f_{m1}\, f_{m2}). \quad \text{(B.6)}$$

Equations (B.3),(B.4), and (B.6) exactly agree in their expression of the Banzhaf index $TBP(X_m)$ for $(1 \leq m \leq n)$. In passing, we note that if the function $f(\boldsymbol{X})$ is a threshold switching function, then its characteristic Chow parameters essentially coincide with the Banzhaf indices $TBP(X_m)$ as given in (5) or (B.4) [130-132].

## Appendix C: Iterative Computation of the Cumulative Combinatorial Coefficient

The binomial (combinatorial) coefficient $c(n,k)$ (defined over $(0 \leq k \leq n)$) satisfies the recursive relation

$$c(n,k) = c(n-1,k) + c(n-1,k-1), \qquad (0 < k < n), \quad \text{(C.1)}$$

together with the boundary conditions

$$c(n,0) = c(n,n) = 1, \quad n \geq 0. \quad \text{(C.2)}$$

We use the upper–case $C(n,k)$ to distinguish the cumulative combinatorial coefficient from the standard combinatorial coefficient depicted by the lower-case $c(n,k)$. We define the cumulative coefficient as

$$C(n,k) = \sum_{m=k}^{n} c(n,m), \qquad (0 \leq k \leq n). \quad \text{(C.3)}$$

This coefficient satisfies a recursive relation similar to that in (C.1), namely

$$C(n,k) = C(n-1,k) + C(n-1,k-1), \qquad (0 < k < n), \quad \text{(C.4)}$$

but with boundary conditions

$$C(n,0) = \sum_{m=0}^{n} c(n,m) = 2^n . \quad \text{(C.5a)}$$

$$C(n,n) = \sum_{m=n}^{n} c(n,m) = c(n,n) = 1. \quad \text{(C.5b)}$$

Note that while the array of $C(n,k)$ values can be obtained from that of $c(n,k)$ values through the summation relation (C.3), the $c(n,k)$ array is obtainable from the $C(n,k)$ array via the knowledge that $c(n,n) = 1$ or $C(n,n+1) = 0$, and the differencing relation

$$c(n,k) = C(n,k) - C(n,k+1). \qquad (0 \leq k < n). \quad \text{(C.6)}$$

Figure C.1 shows Pascal-triangle constructs for $c(n,k)$ (top) and $C(n,k)$ (bottom) on the two dimensional plane of $k$ (vertical downwards) versus $n$ (vertical rightwards). The yellow lines are the boundary lines, namely the top horizontal line $(k = 0)$ and the diagonal line $(k = n)$. Each node within the octant delimited by the two boundaries is obtained by the recursive relation (C.2) for $c(n,k)$ and the recursive relation (C.4) for $C(n,k)$. In both cases, the value of this node is the sum of values of the node to its left and the value of the node on top of this latter node. The cumulative combinatorial coefficient can be efficiently computed through the iterative column-wise construction of the bottom construct in Fig. C.1 up to column $N$. The following Pascal-like fragment achieves this computation.

$C(0,0) \coloneqq 1; \quad C(1,0) \coloneqq 2; \quad C(1,1) \coloneqq 1;$
$For\ n = 2\ to\ N\ do\ begin$
$\quad C(n,n) \coloneqq 1;$
$\quad For\ k = (n-1)\ downto\ 1\ do$
$\qquad C(n,k) = C(n-1,k) + C(n-1,k-1);$
$\quad C(n,0) \coloneqq 2 * C(n-1,0);$
end;

| | | | | | | | | | | |
|---|---|---|---|---|---|---|---|---|---|---|
| 1 | 1 | 1 | 1 | 1 | 1 | 1 | 1 | 1 | 1 | 1 |
| | 1 | 2 | 3 | 4 | 5 | 6 | 7 | 8 | 9 | 10 |
| | | 1 | 3 | 6 | 10 | 15 | 21 | 28 | 36 | 45 |
| | | | 1 | 4 | 10 | 20 | 35 | 56 | 84 | 120 |
| | | | | 1 | 5 | 15 | 35 | 70 | 126 | 210 |
| | | | | | 1 | 6 | 21 | 56 | 126 | 252 |
| | | | | | | 1 | 7 | 28 | 84 | 210 |
| | | | | | | | 1 | 8 | 36 | 120 |
| | | | | | | | | 1 | 9 | 45 |
| | | | | | | | | | 1 | 10 |
| | | | | | | | | | | 1 |
| 1 | 2 | 4 | 8 | 16 | 32 | 64 | 128 | 256 | 512 | 1024 |
| | 1 | 3 | 7 | 15 | 31 | 63 | 127 | 255 | 511 | 1023 |
| | | 1 | 4 | 11 | 26 | 57 | 120 | 247 | 502 | 1013 |
| | | | 1 | 5 | 16 | 42 | 99 | 219 | 466 | 968 |
| | | | | 1 | 6 | 22 | 64 | 163 | 382 | 848 |
| | | | | | 1 | 7 | 29 | 93 | 256 | 638 |
| | | | | | | 1 | 8 | 37 | 130 | 386 |
| | | | | | | | 1 | 9 | 46 | 176 |
| | | | | | | | | 1 | 10 | 56 |
| | | | | | | | | | 1 | 11 |
| | | | | | | | | | | 1 |

Figure C.1. Pascal-triangle constructs for the combinatorial coefficient $c(n,k)$ (top) and the cumulative combinatorial coefficient $C(n,k)$ (bottom) on the two dimensional plane of $k$ (vertical downwards) versus $n$ (vertical rightwards). The green-colored cells are $c(9,3) = 84$ (used in (73) and (104)) and $C(10,4) = 848$ (used in (72) and (103)).

## Appendix D: Preservation of Disjointness When a Boolean Quotient is Constructed

The sum-of-products (s-o-p) form $\bigvee_{k=1}^{m} D_k$ is disjoint if every two products $D_{k_1}$ and $D_{k_2}$ in it are disjoint, i.e., $D_{k_1} \wedge D_{k_2} = 0$. This happens when there is at least one opposition between the two products, i.e., at least one variable $X_j$ that appears complemented in one of these two products and un-complemented in the other [64, 71, 79]. Now if we construct the Boolean quotient of this sum w.r.t. the afore-mentioned literal $X_j$ or w.r.t. its complement (the literal $\bar{X}_j$), then one of these two products diminishes, and hence retains its disjointness with the other. If we construct the Boolean quotient of the sum $\bigvee_{k=1}^{m} D_k$ w.r.t. a literal $X_i$ other than any of the opposition literals $X_j$ or $\bar{X}_j$, then the Boolean quotients $(D_{k_1}/X_i)$ and $(D_{k_2}/X_i)$ remain disjoint, since one of them has $X_j$ as one of its literals, while the other contains $\bar{X}_j$ as one of its literals. This means that disjointness of a sum-of-products form is preserved by the construction of a Boolean quotient.

By contrast, disjointness is not necessarily preserved through partial differentiation. For example, the partial differentiation of the disjoint expression $(P_1 \oplus \bar{P}_1 P_2)$ in (52) w.r.t. $P_1$ results in the non-disjoint expression $(1 \oplus P_2)$ in (53). Likewise, the partial derivative of the disjoint formula of $f(\boldsymbol{X})$ in (B.2) is not disjoint.